\documentstyle[preprint,eqsecnum,psfig,aps]{revtex}

\def\DD{\mbox{\rm D}} 
\def\Pre{\mbox{\rm P}} 
\def\pp{\mbox{\rm p}} 
\def\su{\mbox{\rm U}} 

\input{epsf}
\begin{document}
\draft
\title{On the shear instability of fluid interfaces: \\
Stability analysis
}
\author{A.\ Alexakis}
\address{Department of Physics, University of Chicago \\
Chicago, IL\ \ 60637}
\author{Y.\ Young}
\address{Dept.\ of Engineering Sciences and Applied Mathematics,\\
Northwestern University,
Evanston, IL\ \ 60608}
\author{R. Rosner}
\address{Departments of Astronomy \& Astrophysics and Physics, The
University of Chicago \\
Chicago, IL\ \ 60637}
\date{\today}
\maketitle
\begin{abstract}
We examine the linear stability of fluid interfaces
subjected to a shear flow. Our main object is to generalize previous work
to arbitrary Atwood number, and to allow for surface tension and
weak compressibility.
The motivation derives from instances in astrophysical systems where 
mixing across material interfaces driven by shear flows may 
significantly affect the dynamical evolution of these systems.
\end{abstract}



\section{Introduction}
\label{sec:intro}

The stability of fluid interfaces in the presence of
shear flows has been studied for almost half a century; and was largely
motivated by the problem of accounting for observations of surface water
waves in the presence of winds. As early as the 1950's, it was
realized that classical Kelvin-Helmholtz instability \cite{Chandra62}
could not account for the observed water waves (cf.\
\cite{Miles57,Lighthill62}), and efforts were initiated to study the full
range of possible unstable modes by which interfaces such as represented
by
the water/air interface could become unstable. By the early 1960's, the
basic mechanism was understood, largely on the basis of work by Miles
\cite{Miles59a,Miles59b,Miles62} and Howard \cite{Howard61}: They
discovered that interface waves for which gravity provided the restoring
force (e.g., waves that can be identified with so-called deep water
waves) can be driven unstable via a resonant interaction with the ambient
wind; this work was also one of the first applications in which resonant
(or critical) layers played an essential role in both the physics and the
mathematics. Work carried out at that time showed that the precise form
of the vertical wind shear profile was critical to the nature of the
instability; typically, it was assumed that the wind immediately above
the water surface could be characterized by a logarithmic profile of the
form
\begin{equation}
U(z) = U_o + U_1 \ln (z / \delta +1) ~,
\label{windpr}
\end{equation}
where $U_o$ is the velocity jump (if any) at the water/air interface, $z$
is the vertical coordinate (with $z=0$ marking the initial water-air
interface), and $\delta$ is the characteristic scale length of the shear
flow in the air.\footnote{Such velocity profiles are commonly observed
in the boundary layer of winds blowing over the surface of extensive
bodies of water; cf.\ Miles (1957).} The idea was then to demonstrate
that surface gravity waves whose phase speed is given by
$c=\sqrt{g\lambda A}$ ($g$ the gravitational acceleration, $\lambda$ the
perturbation mode wavelength, and
$A\equiv (\rho_2-\rho_1)/(\rho_2+\rho_1)$ the Atwood number for a density
interface between fluids of density $\rho_1$ [upper fluid] and $\rho_2$
[lower fluid], with $\rho_1 < \rho_2$) can couple to this wind profile at
a height $z$ where $c \sim U(z)$. At the time, it was not possible to
construct a self-consistent description of the problem, such that a
logarithmic wind profile automatically emerged from the analysis; and
much of subsequent work has focused on establishing the nature of this
wind shear profile (e.g., \cite{Miles93}). Finally, we note that these
studies have since been applied to a number of other contexts, including
especially shear flows in atmospheric boundary layers, where they have
been extensively expanded, including into the weakly compressible regime
\cite{Davis76}.

Our own work is originally motivated by an astrophysical problem in
which mixing at a material interface between two fluids with different
densities is essential to the evolution of the astrophysical problem.
Specifically, consider a white dwarf star, whose composition is almost
completely dominated by carbon and oxygen. If such a star is in a close
binary orbit with a normal main sequence star, then it has been known for
some time (e.g., \cite{Starrfield74}) that accretion of matter from the
normal star (largely in the form of hydrogen and helium) can lead to a
build-up of an accretion envelope on the white dwarf which is capable of
initiating nuclear hydrogen ``burning". This burning process can
lead to a nuclear runaway, in which the energy released as a result of
these nuclear fusion reaction is sufficient to expel a large fraction of
the accreted matter in the form of a shell; such a runway is referred to
in the astronomical literature as a ``nova". The key element relevant to
our present discussion is then that observations show that approximately
30\% by mass of the ejecta are in the form of C+O nuclei: since neither
carbon nor oxygen are products of hydrogen burning in the accreted
envelope, it must be the case that some sort of mixing process brought
large amounts of stellar (i.e., white dwarf)
carbon and oxygen into the overlying accreted
material before envelope ejection. Furthermore, detailed analysis of the
energetics of the runaway process has shown that simple hydrogen burning
in the envelope cannot provide enough energy to power the observed nova;
thus, additional energy release via ``catalytic" nuclear reactions in
which C+O play important roles is required in order to match the
observations (cf.\
\cite{Starrfield78a,Starrfield78b,Wallace81,Woosley86,Shankar92,Shankar94}).
Thus, from the perspective of both observed abundances of nova ejecta and
consideration of the nova energetics, efficient mixing at the
star/envelope interface is called for. 
%
%
Several possible mixing processes have been discussed in the literature,
including undershoot driven by thermal convection in the burning envelope 
and Kelvin-Helmholtz instability; but detailed studies shown all of them
to be ineffective in producing the required mixing
(e.g.,\cite{Kercek00a,Kercek00b,KipT78}). In this regard, the current
astrophysical situation resembles the problem encountered by
oceanographers in the 1950's, as they tried to explain the observed mixing
between the sea water-atmosphere interface.  
 A new instability is needed to account for the observed mixing
\cite{Miles57}.

Following the previous oceanographic work, we explore the possibility that
a critical-layer instability related to the coupling of stellar surface
gravity waves to a shear flow in the hydrogen envelope - can account for
the enhanced mixing rate. Thus, in this paper we embark on a systematic
study of such an instability  and apply our results to the specific case
of mixing of C/O to H/He envelope of white dwarf stars \cite{Rosner01}.
 We note that
similar scenarios can arise in a variety of other astrophysical systems,
such as in the boundary layer between an accretion disk and a compact
star, where mixing between fluids of different densities -- as in the nova
problem 
--
is expected to play an important role. However, the earlier non
astrophysical work largely focused on the case of very large density
differences between
the two fluids separated by an interface, and primarily considered the
fully incompressible case (the weakly compressible case has been
considered by \cite{Davis76}). For the astrophysical case, the density
ratios can be of order unity (for the nova case a typical value would be
$\rho_1/\rho_2 \sim 1/10$)
 and the Mach number for the interface between the 
accretion flow and the white dwarfs surface can range from very subsonic
to of order 0.2.
 The aim of this paper
therefore is to extend the shear flow
analysis to arbitrary density ratios , shear and compressibility.
 We provide estimates of the growth rates of
unstable surface waves, and determine the regions in the control
 parameter space that correspond to different instabilities for different
physical situations.

This paper is structured as follows: In the next section we define the
problem to be solved more precisely; \S \ref{sec:linear} describes the
linear analysis
for the incompressible, two-layer case. \S \ref{sec:tension} and \S
\ref{sec:compress} 
describe, respectively, the inclusion of surface tension and extension
to compressible flow of low Mach numbers.
We discuss and summarize our results in the final section.


\section{Formulation}
\label{sec:formulation}
The starting point of our formulation is the identification of the
appropriate material equations of motion. This issue has been
well-discussed in the literature, including the motivating white dwarf
case \cite{stelar68}: in general, we can expect the gaseous surface and
atmosphere of such stars to be well described by the single fluid
equations for an ideal gas. More specifically, the length scales of the
physical mixing processes discussed here are all far larger than the Debye
length, so that the ionized stellar material can be considered to be
neutral;  and as long as the stellar magnetic fields are weak (e.g.,
$\beta \equiv$ gas preassure/magnetic preassure $\gg 1$) we can ignore the
Lorentz force. Furthermore, the ratio of the spatial scales of interest to
the Kolmogorov scale is large (typically $ > 10^4$) so that we are in the
large Reynolds number limit, and viscous effects on the motions of
interest will be negligible.  As a consequence, the Euler equations will
be describing our system.
 
 We consider a two-dimensional
flow with $x$ the horizontal direction and $y$ the vertical.
The system consists a layer of light fluid (density $\rho_1$)
on top of a layer of heavy fluid (density $\rho_2$).
In most of our analysis $\rho_1$ and $\rho_2$ are constant in each layer,
and in the most
general scenario both layers can be stratified
(densities are functions of $y$).
The two layers are separated by an interface
given by $y=h(x;t)$, which initially is
taken to be flat ($y=h(x;0)=0$).
The upper layer ($\rho_1$) is moving with
velocity $\su(y)$ in the $x$ direction parallel
to the initial flat interface,
while the lower
layer ($\rho_2$) remains still.

As already mentioned, the instability of such stratified shear flow
has been investigated (cf. \cite{Miles57,Howard61}), albeit under
limited physical circumstances. We study this problem in
full generality, allowing for a variety of effects (including broad
ranges in the values of the Atwood number/gravity and in
compressibility) with the motivation that one can establish the role of the
relevant instabilities under more general astrophysical circumstances than
the restricted case of nova-related mixing which motivated our study.


\subsection{The general problem}
\label{subsec:1.1}

A wind (shear flow) is assumed to flow only in the layer of light
fluid ($\rho_1$) and is zero in the heavy fluid ($\rho_2$). Within each
layer, the governing equations are the continuity equation

\begin{equation}
\label{eq_continuity}
\partial_t \rho + \nabla\cdot (\rho \vec{u} )=0,
\end{equation}
and the two-dimensional Euler equation

\begin{equation}
\label{eq_Euler}
\rho \partial_t
\vec{u} + \rho \vec{u}\cdot\nabla \vec{u} = -\nabla P +\rho \vec{g}.
\end{equation}
The equation of state closes the system, which is expressed in dynamical
terms:

\begin{equation} \label{eq_eos} (\partial_t + \vec{u}\cdot\nabla) \Pre =
\frac{\DD \Pre}{\DD t} = \frac{\gamma\Pre}{\rho} \frac{\DD\rho}{\DD t},
\end{equation}

\noindent
where $\gamma$ is the polytropic exponent. The background density and
pressure are in hydrostatic equilibrium, $\partial_yP_o=-\rho_o g$. The
basic state is then defined by a shear flow $(\su(y))$ in the upper
layer, and hydrostatic pressure $(P_o)$ and density profiles $(\rho_o)$.
We perturb around this basic state
\begin{equation}
\label{pertub_var}
\vec{u} = \su(y)\hat{x} + \vec{u^{\prime}},\;\;\;
\rho = \rho_0(y) + \rho^{\prime},\;\;\;
\Pre = \Pre_0(y) + \pp^{\prime},
\end{equation}
and study the growth of the perturbations (primed variables). From
equation \ref{eq_eos}, the density and pressure perturbations satisfy the
relation

\begin{equation}
\label{delta_pres_dens}
\frac{\DD \pp^{\prime}}{\DD t} = c_s^2\frac{\DD\rho^{\prime}}{\DD t} +
{w^{\prime}(g\rho_0 + c_s^2\rho_{0y}}),
\end{equation}
where $w^{\prime}$ is the vertical component of the perturbation velocity,
$g$ is the gravitational acceleration, and $c_s=\sqrt{\gamma P_0/\rho_0}$
is the sound speed for the background state. Upon expanding the
perturbations in normal modes $e^{ik(x-ct)}$, we obtain the linearized
equations in the perturbation quantities (where we have dropped the
primes for convenience):

\begin{equation}
\label{lin_comp_1}
\begin{array}{lll}
ik(\su-c)u &+w\partial_yU &= -ik \rho_o^{-1} p \,, \\
ik(\su-c)w & &= -\rho_o^{-1} \partial_yp - \rho_o^{-1}g \rho \,, \\
ik(\su-c)\rho& & = -\rho_o(iku+\partial_yw)-w\partial_y\rho_o \,, \\
ik(\su-c)p &-wg\rho_o &= c_s^2
\left( ik(U-c)\rho+w\partial_y\rho_o \right) \,.
\end{array}
\end{equation}

The above equations form an eigenvalue problem for the complex number
$c$. One immediately sees that the incompressibility condition
$\nabla\cdot \vec{u}=0$ can be obtained by taking the limit $c_s \to
\infty$. Our problem simplifies greatly with this assumption.
Therefore we first present our results for the incompressible case, and
then examine how compressibility modifies the stability properties.


\subsection{The incompressible case}
\label{subsec:1.2}

For the incompressible case we define a stream function $\Psi$ such that
$u=\partial_y \Psi$ and $w=-\partial_x\Psi$. The 2-D Euler equation thus
reads

\begin{equation}
\partial_t \nabla^2\Psi-\Psi_x\nabla^2\Psi_y+\Psi_y\nabla^2\Psi_x
=0, \quad y\ne h.
\end{equation}
The total stream function $\Psi=\Psi_0+\psi$ consists of a
background stream function $\Psi_0=
\int_0^y\su(z)dz$ and a perturbation $\psi=\phi(y)
e^{ik(x-ct)}$.
The linear equation for $\phi$ is the well-studied Rayleigh equation,

\begin{equation}
\label{linear_Rayleigh}
\phi ^{\prime\prime}- \Big{(}
k^2 + \frac{\su^{\prime\prime}}{\su-c} \Big{)}\phi=0.
\end{equation}
The boundary conditions at the interface for the continuity
of the normal component of the velocity and pressure are:

\begin{eqnarray}
\label{linear_bc1}
  (\su-c) \tilde{h} -\phi^{\pm} &=& 0 ,\;\;\;\\
\label{linear_bc2}
  \Delta \big{[} \rho_i( (\su-c)\phi^{\prime} -\su^{\prime} \phi) \big{]}
  +g\tilde{h}(\rho_1-\rho_2)&=&0 ,
\end{eqnarray}

\noindent
where $\Delta$ indicates the difference across the interface, and
$\tilde{h}$ is the amplitude of the perturbed interface,
$h=\tilde{h}e^{ik(x-ct)}$.


\subsection{The compressible case}
\label{subsec:1.3}

For the compressible case, where $c_s$ is comparable to the background
shear flow and the density stratification is non-negligible on the scales
of interests, we start from the full set of equations \ref{lin_comp_1}.
We obtain the following equations by eliminating $\rho$ and $u$:
\begin{equation}
\label{eq:linear_comp}
\begin{array}{ll}
\rho_o(k^2U_G^2+gk_g+gk_s)w &=ikU_G(\partial_y+k_g)p \,, \\
ik^2(U_G^2/c_s^2-1)p &= \rho_ok(k_gU_G+\partial_yU_G-U_G\partial_y) w \,,
\end{array}
\end{equation}
where $U_G=U-c$ is the Galilean-transformed velocity in the reference
frame of the wave, $k_s=\rho_o^{-1} \partial_y \rho_o$ is the inverse
stratification length scale, and $k_g=g/c_s^2$. We further simplify the
equations by applying the transformation
\cite{Chimonas}
\[p=f^{-1}\tilde{p}, \; \;\;w=iU_Gqf,\;\;\;\tilde{\rho}_o=\rho_of^2\;\;\;
\mbox{with} \;\;
f=e^{\int_0^yk_g(z)dz}. \]
Equations \ref{eq:linear_comp} are then rewritten in terms of these new
variables as follows:

\begin{equation}
\begin{array}{ll}
\tilde{\rho}(k^2U_G^2+gk_g+gk_s)q &=k\partial_y \tilde{p}, \\
k^2 \left(1-U_G^2/c_s^2 \right)\tilde{p}
  &=\tilde{\rho}_ok U_G^2\partial_y q,
\end{array}
\end{equation}
which can be combined to give \cite{Davis76}

\begin{equation}
\label{q_comp_eq}
\partial_y
\left(\frac{\tilde{\rho}_oU_G^2\partial_yq}{1-U_G^2/c_s^2}
\right)
-\tilde{\rho}_o\Big{[}k^2U_G^2+g(k_s+k_g)\Big{]}q=0 \,.
\end{equation}

\noindent
We re-write the above equation into canonical form.
The resulting equation is similar to the
Rayleigh equation for the incompressible flow, except for an additional
stratification term $-g(k_s+k_g)/\tilde{\su}^2_G\phi$:

\begin{equation}
\label{comp_1}
\partial_y^2\phi-\left[\kappa^2+g\tilde{\rho}_o
\frac{k_s+k_g}{\tilde{\su}_G^2}+
\frac{\partial_y^2\tilde{\su}_G}{\tilde{\su}_G} \right]\phi=0
\end{equation}

\noindent
where $\kappa^2=k^2\left(1-U_G^2/c_s^2\right)$,
$\tilde{\su}_G=kU_G\sqrt{\tilde{\rho}_o}/ \kappa$ and
$\phi=q\tilde{\su}_G/k=-ik^{-1}
\sqrt{\rho_o}w\left(1-U_G^2/c_s^2\right)^{-1}$.
It can be shown that the stratified Rayleigh equation can be recovered by
taking the limit of $c_s \to \infty$.
\begin{equation}
\label{strat_1}
\partial_y^2\phi-\left[k^2+g\frac{k_s}{\su_G^2}+
\frac{\partial_y^2(\su_G\sqrt{\rho_o})}{\su_G\sqrt{\rho_o}} \right]\phi=0
\end{equation}
Furthermore, we recover the unstratified Rayleigh equation 
in the same limit, if $k_s+k_g=0$
(which corresponds to an adiabatic atmosphere, as we will show later on).
Finally, the boundary conditions at the interface are expressed in terms
of $\tilde{\su}_G$ and $\phi$:
\begin{equation}
\label{comp_bc1}
q=\frac{\phi^+}{\su_G^+}=\frac{\phi^-}{\su_G^-}=\tilde{h}
\end{equation}
using the continuity of $q$ and integrating \ref{q_comp_eq} across the 
interface we obtain
\begin{equation}
\label{comp_bc2}
\Delta
\Big{[}\tilde{U}_G\partial_y\phi
-(\partial_y\tilde{U}_{G})\phi
-g\tilde{\rho}\phi/\tilde{\su}_G\Big{]}=0 \,.
\end{equation}


\subsection{Wind profiles}
\label{subsec:1.4}

In general, it is not trivial to determine the wind profile: strictly
speaking, the wind profile should be determined as part of the solution
of the evolution equation for the wind shear interface. However, it has
been customary to simplify the problem by assuming an a priori analytical
form for the wind profile, and to use this in order to study the stability
properties of the interface; thus, Miles \cite{Miles57} used a
logarithmic wind profile from turbulent boundary layer theory to model
the wind profile in the air over the ocean. In this example, the
turbulence level in the wind is simply defined by the scale height of the
wind profile, which in turn simply depends on how ``rough" the boundary
is.

In our formulation we shall also assume the wind profile to be of simple
form, and scale distance with respect to the length scale of the wind
boundary layer. In order to explore the sensitivity of our results to
the nature of this wind boundary layer, we will examine two different
kinds of wind profiles: the first is the logarithmic wind profile
$\su(y)=U_0+U_1\ln(ay+1)$, which is derived from turbulent boundary layer
theory for the average flow above the sea surface; the second is given by
$\su(y)=U_1 \tanh (ay)$, which has the more realistic feature of
reaching a constant finite flow speed above the interface.


\section{Linear analysis: incompressible case}   
\label{sec:linear}

We start with the stability analysis of the incompressible case
with constant densities in the two layers.
The fluid is described by the Rayleigh equation
\ref{linear_Rayleigh} within each layer; and we ignore surface tension
for the time being. We solve the following equation in each layer:

\begin{equation}
\label{linear_Rayleigh_2}
\phi_{yy} - \Big{(} {k}^2 +\frac{U_{yy}}{U-c}
\Big{)}\phi=0,\quad \quad \phi|_{y \to \pm \infty}=0,
\end{equation}
with boundary condition (at $y=0$)

\begin{equation}
\rho_2kc^2-\rho_1\Big{[}(U-c)^2 \phi_y-(U-c)U_y
\Big{]}-g(\rho_2-\rho_1)=0,
\end{equation}
where we have normalized $\phi$ by setting $ \phi|_{y=0}=1 $.

We scale lengths by $a^{-1}$, the characteristic length of the wind
profile \footnote{In oceanography, such a length scale is
referred to as the ``roughness" of the wind profile.} and the velocity by
the reference velocity $U_1$. The dimensionless equation thus reads

\begin{equation}
\label{linear_rescale_01}
\phi_{yy} - \Big{(}
{K}^2 +\frac{V_{yy}}{V-C} \Big{)}\phi=0,\quad \phi|_{y=0}=1,\quad
\phi|_{y=\infty}=0;
\end{equation}
and the boundary condition at the interface now becomes

\begin{equation}
\label{linear_rescale_02}
KC^2-r\Big{[}(V_0-C)^2\phi_y-(V_0-C)V_y \Big{]}-G(1-r)=0.
\end{equation}
where $C=c/U_1$,
$K=k/a$, $G=g/\su_1^2a$,
$V_0=\su(0)/\su_1$, and $r=\rho_1/\rho_2$.

For a given wind profile, the system then is characterized by the four
parameters $(K, G, V_0, r)$. Parameter $G$ measures the ratio of
potential energy associated with the surface wave to the kinetic energy
in the wind. The Richardson number defined in stratified
shear flow is not useful in quantifying the stability in our case. However,
as will be shown later,
we find parameter $G$ to be a good substitute in
describing the effect of stratification on the surface wave instability.
 In the case of accretion flow on the surface of a white
dwarf $G\sim 1$, while in the case of oceanic waves driven by winds,
$0.1<G<1.0$. Table \ref{table1} lists the values of $G$ for a variety of
physical conditions.

The aim of our linear analysis then is to find the value of $C$ in the
complex plane as a function of these 4 parameters, and to establish the
stability boundaries in the space $(K,G,r,V_o)$; note that in our
convention, $\Im \{C\}> 0$ implies instability
(where $\Im \{ \}$ refers to taking the imaginary part).


\subsection{Kelvin-Helmholtz modes and Critical Layer modes}
\label{subsec:KH_vs_CL}

Before solving this problem, some general remarks about the set of
equations \ref{linear_rescale_01} - \ref{linear_rescale_02} are
required. We observe that in the inviscid limit, if $C$ is an eigenvalue,
then so is $C^*$; therefore we will have a stable wave only if $\Im \{C\}=0$.
If that is the case, then at the height where
$V_c\equiv V(y_{cr})=C$ (assuming such a height exists) the Rayleigh
equation has a singularity; this location $y=y_{cr}$ is called the
critical layer, and is well-discussed in the literature
\cite{Drazin81,Lin55}

The existence of such a critical layer is crucial for the presense of
instability. One can prove (Appendix A) that our system can be
unstable only if $C_r\equiv \Re\{C\} \le V_{max}$.
For the case that $V_o=0$ there always exists a point in the flow where
$C_r=V \le V_{max}$ for all unstable modes.
We denote this point as a critical layer even if $C$ is complex i.e.,
$C_i\equiv \Im\{C\} \not= 0$; and thus there is no singularity. However,
if $V_o\ne 0$, such a point might not exist (e.g. if $C_r<V_o$). In that
case the only mechanism that can destabilize the flow would be a
Kelvin-Helmholtz instability. These two kinds of instabilities exhibit
very different properties, both in terms of the physical mechanisms
involved as well as in the mathematical treatment required. Hence we need
to distinguish between (i) modes becoming unstable due to the
discontinuity of the wind profile (from now on called KH-modes), and (ii)
modes becoming unstable due the presense of a critical layer (from now on
called CL-modes).

The KH-modes have been studied for over a century; here we summarize
the results for a
step function wind profile and
some of the features can 
also found for other wind profiles \cite{Chandra62}.
The dispersion relation is given by
\begin{equation}
\rho_2c^2+\rho_1(c+U)^2=(\rho_2-\rho_1)g/k,
\end{equation}
where $U$ is the jump in velocity across the interface between $\rho_1$ and
$\rho_2$. The pressure perturbation $\rho_1(c+U)^2$, providing the
driving force for the instability, is always in phase with the wave, and
is independent of the wavelength. The restoring force
$(\rho_2-\rho_1)g/k$ on the other hand is proportional to the wavelength,
and so we have instability when the wavelength is sufficiently small for
the pressure to overcome the restoring force. In more physical terms,
the flow stream lines above the crests of the perturbed interface wave are
compressed, and above the troughs are decompressed. According to
Bernoulli's equation the pressure above the crests is therefore decreased,
and is increased above the troughs. The wave thus becomes unstable when
these destabilizing pressure forces exceed the stabilizing effects of
gravity. The dispersion relation
\begin{equation}
c=\frac{\rho_1}{\rho_1+\rho_2}U\pm
\sqrt{\frac{g}{k}\left(\frac{\rho_2-\rho_1}{\rho_1+\rho_2}\right) -
\frac{\rho_1\rho_2}{(\rho_1+\rho_2)^2}U^2},
\end{equation}
also shows that the growth rate becomes positive only for wavenumbers
$k>g(\rho_2^2-\rho_1^2)/(U^2\rho_1\rho_2)$.

The CL-modes behave very differently. The solutions of
\ref{linear_rescale_01} near the critical layer for small or zero $C_i$
have a singular behavior. The two Frobenius solutions at the point where
$y=y_{cr}$ are given by
\begin{equation}
\phi_a=z+\left(\frac{\partial_y^2 V}{2\partial_y V}\right)_{cr}z^2+... \; ,
\end{equation}
\begin{equation}
\phi_b=1+\left(\frac{k^2}{2}+\frac{\partial_y^3 V}{2\partial_y V}
+\frac{(\partial_y^2 V)^2}{(\partial_y V)^2}
\right)_{cr}z^2+\dots +
\left(\frac{\partial_y^2 V}{\partial_y V}\right)_{cr}\phi_a(z)\ln|z| ~,
\end{equation}
where $z=y-y_{cr}$ (subscript $cr$ means ``evaluated at the critical
point"). The singular behavior appears in the first derivative
of $\phi_b$. The singularity is removed either because $C_i\not= 0$, in
which case the Frobenius solutions have the same form but $y_{cr}$ is now
complex (so $z$ never becomes zero); or because viscosity becomes
important in this narrow region, in which case the inner solution can be
expressed in terms of generalized Airy functions \cite{Drazin81}. In
either case, the basic result is that there is a phase change across the
critical layer, by which we mean that if $\; \phi=a\phi_a+b\phi_b \;$
is the solution for the stream function above the critical layer, then the
solution below would be $\phi=(a+i\pi b)\phi_a+b\phi_b$ in the previous
formula. Physically this means that the perturbation wave above the
critical layer is not in phase with the wave below this layer. Moreover,
when we apply the boundary conditions at the interface, since
$\partial_y\phi|_0$ is now in general complex, the solution
of equation \ref{linear_rescale_02} will give a complex value of $C$.
That is, the pressure gradient reaches minimum value not on top of the
crests, but rather in front of the crests, where gravity does not act as
effectively as a restoring force. In particular, the destabilizing force
is now non-zero at the nodes of the boundary displacement field (i.e.,
where $h=0$), where the gravitational restoring force vanishes, but
where the vertical velocity of the interface is maximum; thus, the forcing
resembles pushing a pendulum at its point of maximum velocity but minimum
displacement. Note that in this case, there is thus no cut-off for
CL-modes corresponding to the wavenumber cut-off due to gravity for
KH-modes.

Having discussed the physical mechanisms for destabilization, we now turn
to the implications for our choices of initial wind profiles. For wind
profiles with $V_o=0$ one notices that if we assume
$\phi_y$ to be real and known, then the complex eigenvalue $C$ is
obtained by solving equation \ref{linear_rescale_02}
\[C=\frac{r\pm\sqrt{r^2+4G(1-r)(K-r\phi_y)}}{2(K-r\phi_y)}, \]
which will have a nonzero imaginary component only if $\phi_y$ is
positive and $(K-r\phi_y)<0$. However, the negative real part of $C$
implies that the surface wave would be traveling in the direction
opposite to that of the wind --- this case can be excluded on physical
grounds (a more rigorous proof is given in appendix A where we show
 that $C_r>0$).
 Thus, the mechanism that gives rise to the unstable KH-modes
can be excluded. Thus we conclude that surface waves become unstable in
this case only if a critical layer exists. If we use the logarithmic wind
profile, we obtain unstable waves for all wavenumbers because
$\ln(y+1)$ is an unbounded function, therefore a point $y$ where
$C_r=V(y)$ exists for every value of $C_r$. This however is not true for
the tanh wind profile. Because waves with $C_r>V_{max}$ are stable and
$C_r$ (in the absence of surface tension) is a decreasing function of
$K$, there must be a lower bound on $K$, $K_{min}$, so that waves with
$K<K_{min}$ are stable, and unstable otherwise. The value of $K_{min}$ in 
general will depend on the exact form of the wind profile. In appendix B we 
find the exact value of $K_{min}$ for a wind profile of the form 
$V=1-e^{-y}$,
\begin{equation}
K_{min}=\frac{G(1-r)+r-r\sqrt{(G(1-r)+r)^2+(1-r^2)}}{1-r^2} \;\; .
\label{bound1}
\end{equation}  
We remark the following about the previous formula. First of all we 
note that although the previous result holds only for the specific
wind profile used, it can provide a general estimate of $K_{min}$.
 Moreover we note that, unlike the Kelvin-Helmholtz case, in the limit
$r \to 0$, $K_{min}$ remains finite and equal to $G$ (however, 
the growth rate goes to 0 linearly with $r$ i.e., $KC_i\sim r$); this confirms
that for small density ratios CL-modes dominate.
Finally by writing the
wind profile in its dimensional form $U=U_1(1-e^{-ay})$ and taking the
limit  $a \to \infty$ (which takes the wind profile to the 
limiting form of a step-function,
$U=U_1$ for $y>0$ and $U_1=0$ otherwise) we get $ k_{min}=g(1-r)/U^2 $
 which is different from
the result Kelvin-Helmholtz instability gives. We therefore conclude
that different limiting procedures lead to different results.

The situation is more complicated if $V_o > 0$. No critical layer
exists for 
$C_r$ smaller than $V_o$.
Hence, modes of
sufficiently large $K$ become stable (e.g., there is a upper bound on $K$
for the unstable CL-modes). One notes that $C=V_o=\sqrt{G(1-r)/K}$ is a
solution to equation \ref{linear_rescale_02}. This solution
corresponds to the case where the critical layer is right at the
interface. For slower modes than this ($C_r<V_o$) a critical layer will
not exist, and therefore the surface gravity modes will be stable. Thus
CL-unstable modes exist only for $K<G(1-r)/V_o^2$; this result has been
confirmed numerically. As $K$ is increased, the discontinuity of the wind
profile becomes important and Kelvin-Helmholtz instability rises. The
system therefore will be unstable for $K<K_{CL}$ and for $K>K_{KH}$, where
$K_{CL}$ is the upper bound of the CL-modes and $K_{KH}$ is the lower
bound for the KH-modes. This implies that there is a band of
wave numbers $K_{CL}<K<K_{KH}$ that corresponds to stable modes, and
separates the two unstable wavenumber domains. However, as we
will show later, for some values of the control parameters this stable
region disappears, and the two instabilities overlap.

\subsection{Small density ratio}
\label{subsec:small_r}

We are now ready to present results from the linear analysis
for the logarithmic and the tanh wind profiles. The existing literature
has primarily covered the case of small $r$, with the other parameters
assumed to be of order one. In contrast, we are interested in covering a
wider range of the control parameters, and thus provide a complete
description of the full dispersion relation $C = C(K)$. We therefore
briefly summarize Miles' results and move on to the general case.

Assuming the mass density ratio $r$ is a small number (which is true for
the air over water case) and the other parameters are of order one,
Miles \cite{Miles57} expanded the eigenfunction and the wave velocity $C$
with respect to $r$
\begin{eqnarray}
\label{small_r_01}
\phi = r \phi_0 +r^2 \phi_1+ r^3\phi_2 + h.o.t.,\;\;&&
C = C_0+r C_1+ r^2 C_2 + h.o.t. \; ;
\end{eqnarray}
one then obtains the zeroth order solution as a linear gravity wave
with constant amplitude and phase speed $C_0=\sqrt{G/K}$. At first order
${\cal O}(r)$, one finds
\begin{eqnarray}
\label{small_r_02}
\partial_y^2 \phi- \left( K^2 +\frac{\partial_y^2 V}{V_0-C_0} \right)
\phi_0&=&0, \\
\label{small_r_03}
2KC_0C_1-(V_0-C_0)^2 \partial_y \phi_{0}+(V_0-C_0)+G &=&0.
\end{eqnarray}
The asymptotic expansion breaks down at the critical point $y=y_{cr}$
since to first order $C_0$ is real. Using the phase change of $i\pi$ rule
across the critical layer from theory \cite{Drazin81}, Miles obtains the
growth rate of the perturbation at leading order in $r$:
\begin{equation}
\label{small_r_04}
  \Im(C_1)=\frac{1}{2K}(V_0-C_0)^2 
\Im\{\partial_y\phi_{0}\}=-\pi
\frac{(V_0-C_0)^2}{2K}
  \left(\frac{\partial_y^2V}{\partial_yV}\right)_{cr} | \phi_{cr} |^2,
\end{equation}
where the last relation is obtained by multiplying equation
\ref{linear_rescale_01} with the complex conjugate of $\phi$ and taking
the principal value integral, with the contour going below the
singularity; the subscript ``cr" means evaluated at the critical
point.

\begin{figure}[htbp]
\epsfysize=3.0in
\epsfxsize=4.0in
\epsfbox {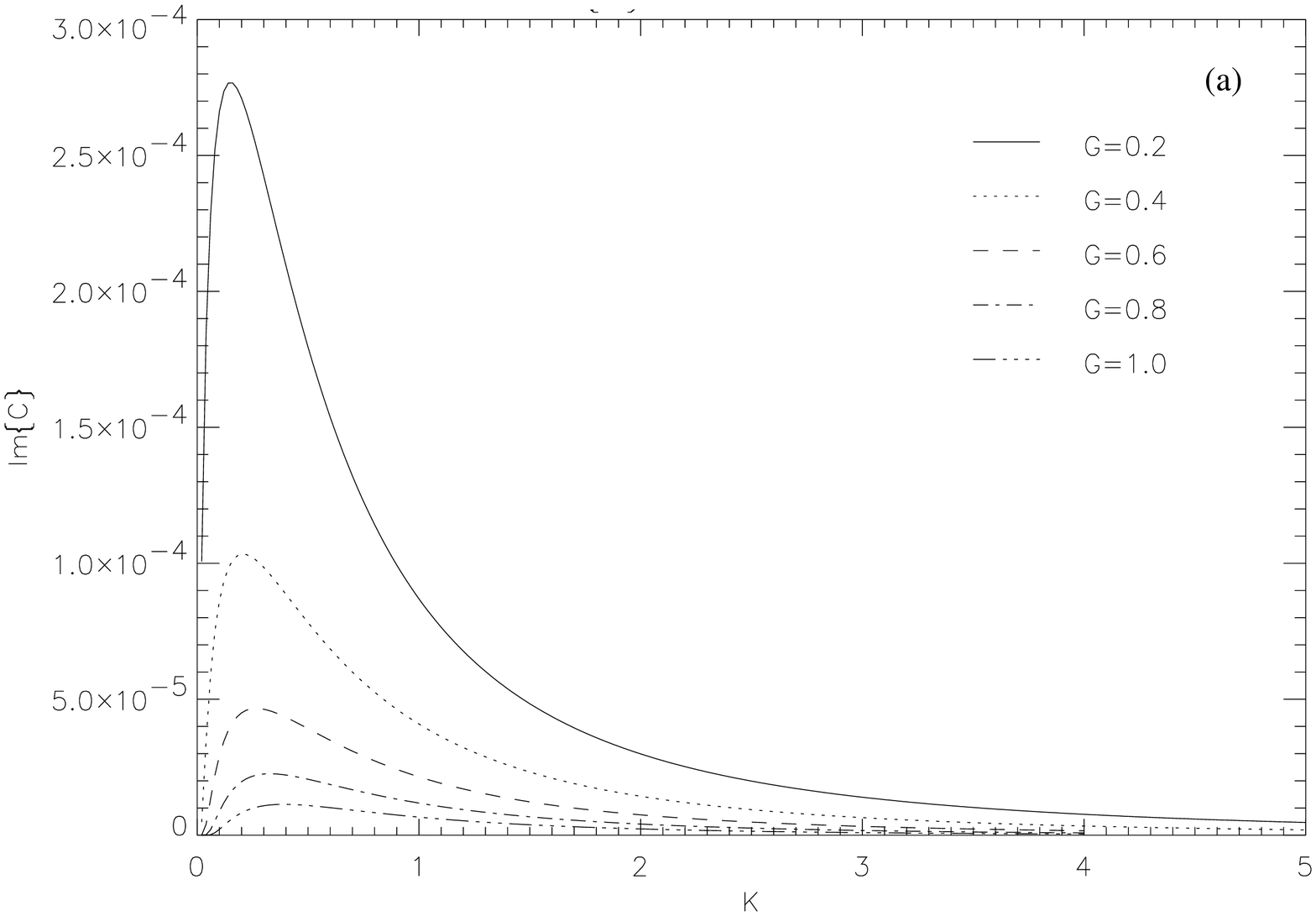}
\epsfysize=3.0in
\epsfxsize=4.0in
\epsfbox {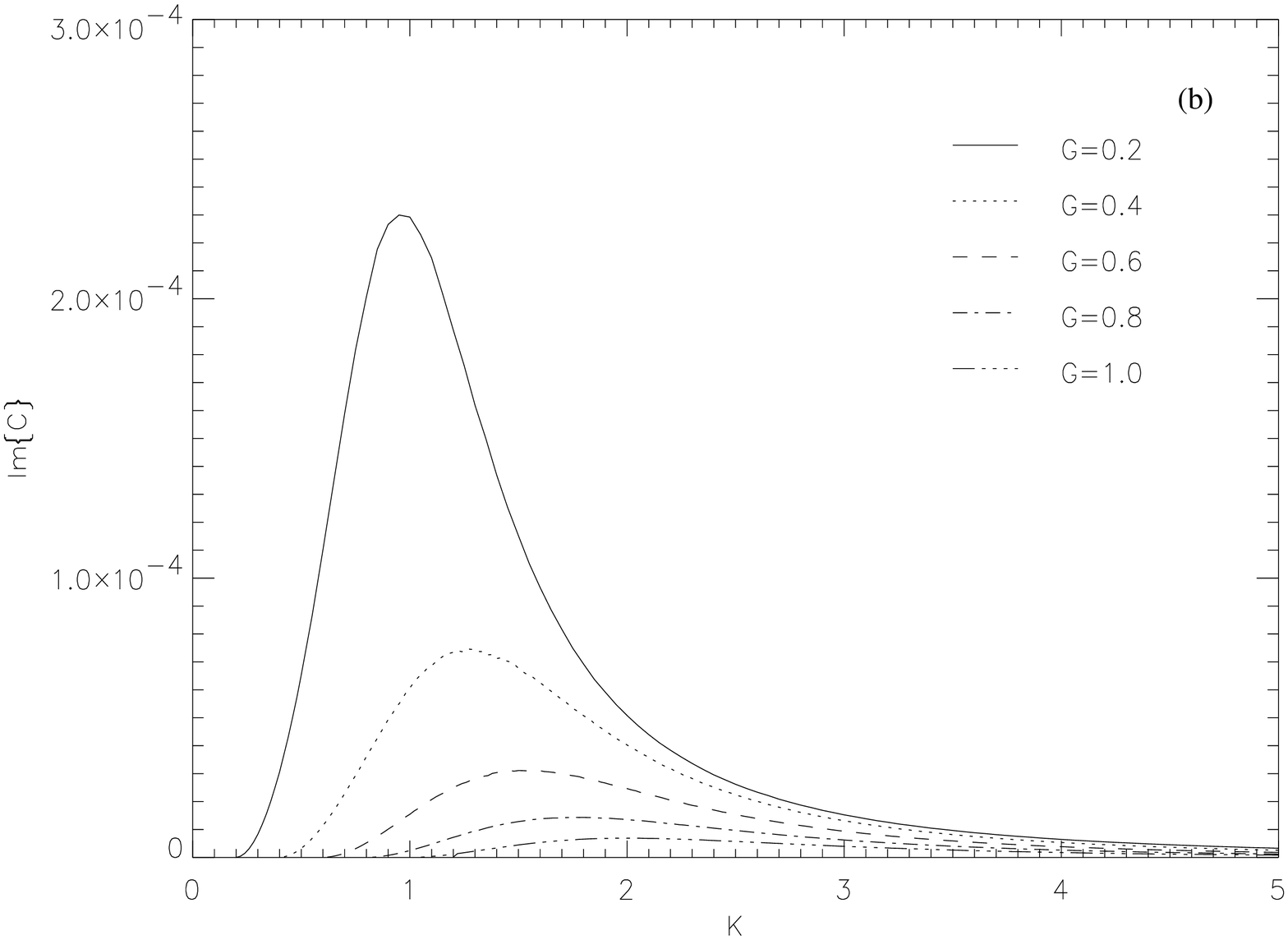}
\caption{Imaginary part of $C$ for $r=0.001$
(a)logarithmic wind profile, (b)tanh wind profile.
Note that the growth rate is given by $K\Im\{C\}$ .}
\label{fig1}
\end{figure}

The first case we examine is when the velocity at the interface is zero.
This simplifies things slightly because, as we discussed before, there are
no Kelvin-Helmholtz unstable modes in this case. The dispersion relation
$\Im(C(K))$ is shown in figure \ref{fig1}a,b for the logarithmic
and for the tanh wind profile for various values of
G. The only difference between the two wind profiles appears at
small wavenumbers: the tanh wind profile (whose asymptotic wind speed is
bounded) does not permit waves traveling faster than the wind to become
unstable. For this reason there is a cut-off which can be found in our
small $r$ approximation to be at $K=G$ for the tanh wind profile.

Next we look at the case where $V_0 > 0$. Now we have both modes present.
As discussed before, the CL-modes are stable for wavenumber $K\le
K_{CL}=G(1-r)/V_o^2$. The KH-modes will appear when $V_oK/G$ increases to
order $1/r$. If we denote by $K_{KH}$ the minimum value of $K$ that the KH
instability is allowed, then for the KH modes in the small $r$ limit $K$
scales as $1/r$, and one can perform a regular perturbation expansion for
large $K$, small $r$ (Appendix C) to find:

\begin{equation}
\label{expansion}
\phi_y|_o=-K-\frac{1}{2}K^{-1}\frac{\partial_y^2V}{V_o -C}-\frac{1}{4}
\Big{[}\frac{\partial_y^3V_o}{V_o-C}-
\frac{\partial_y^2 V_o \partial_y V_o}{(V_o-C)^2}
\Big{]}K^{-2}+... ~,
\end{equation}

\[C=\sqrt{r}\sqrt{\frac{G}{rK}-V_o^2}+rV_o+... ~, \]
and

\[K_{KH}=\frac{G}{Vo^2} \frac{1}{r}-(V_oV'|_o+G)+... ~. \]

The above resembles result for a step function wind profile except for small
corrections due to the non-constant velocity profile. Thus
for small density ratio, the difference between the two modes 
is as discussed in \S \ref{subsec:KH_vs_CL}.
We will discuss the
two instabilities in more detail in the $r={\cal O}(1)$ case.


\subsection{Large density ratio}
\label{subsec:large_r}

\begin{figure}[htbp]
\epsfysize=2.5in
\epsfxsize=4.0in
\epsfbox{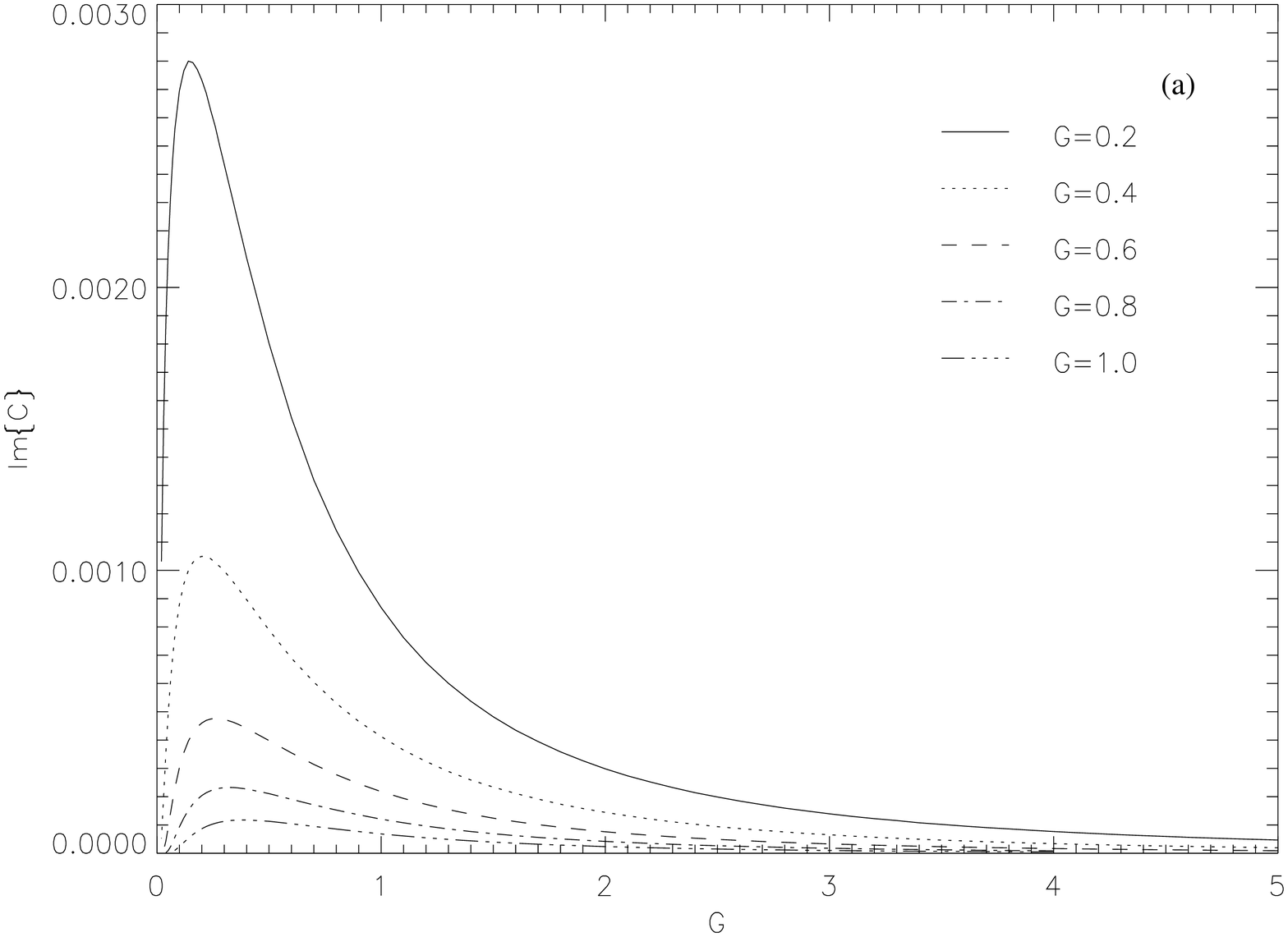}
\epsfysize=2.5in
\epsfxsize=4.0in
\epsfbox{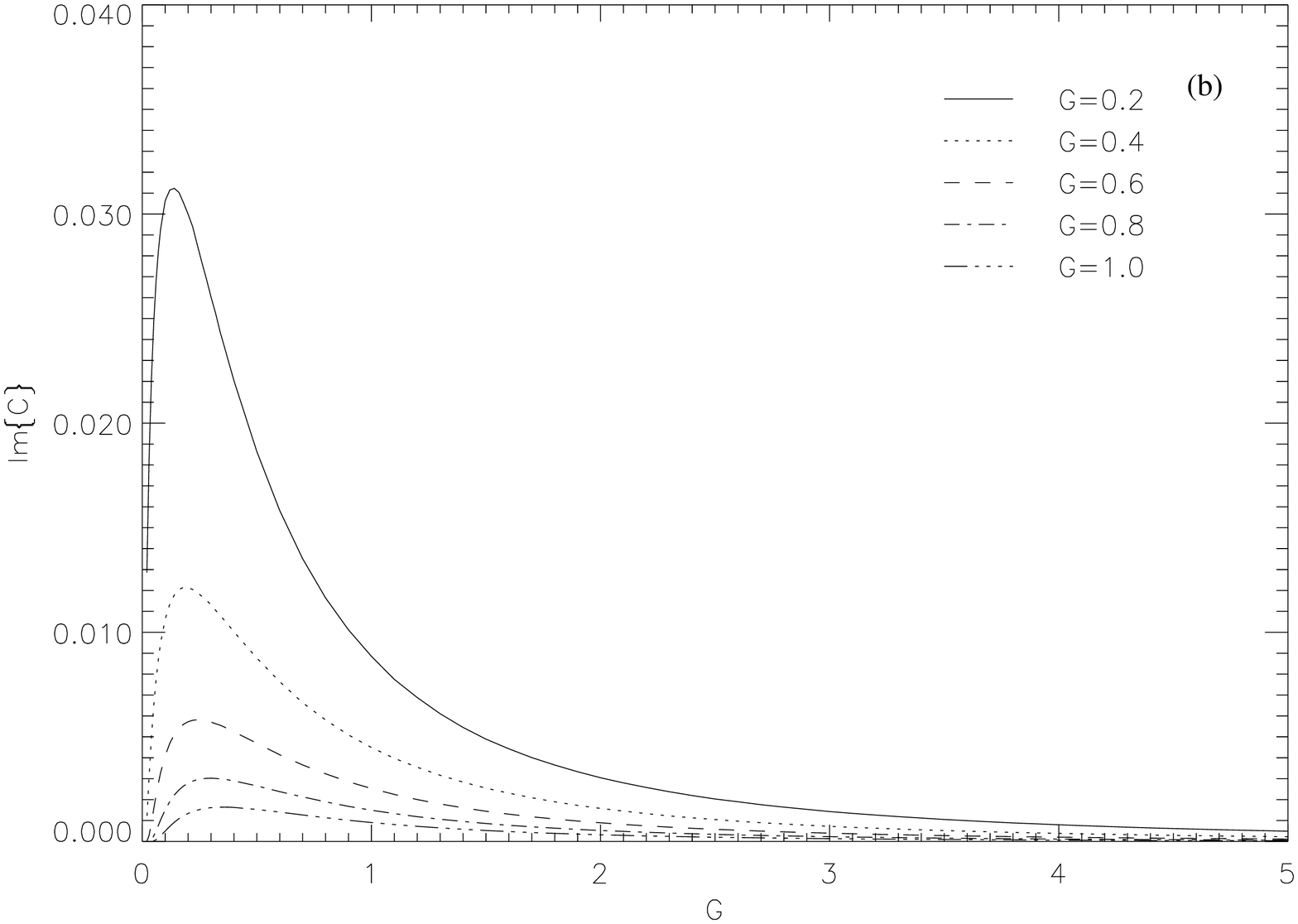}
\epsfysize=2.5in
\epsfxsize=4.0in
\epsfbox{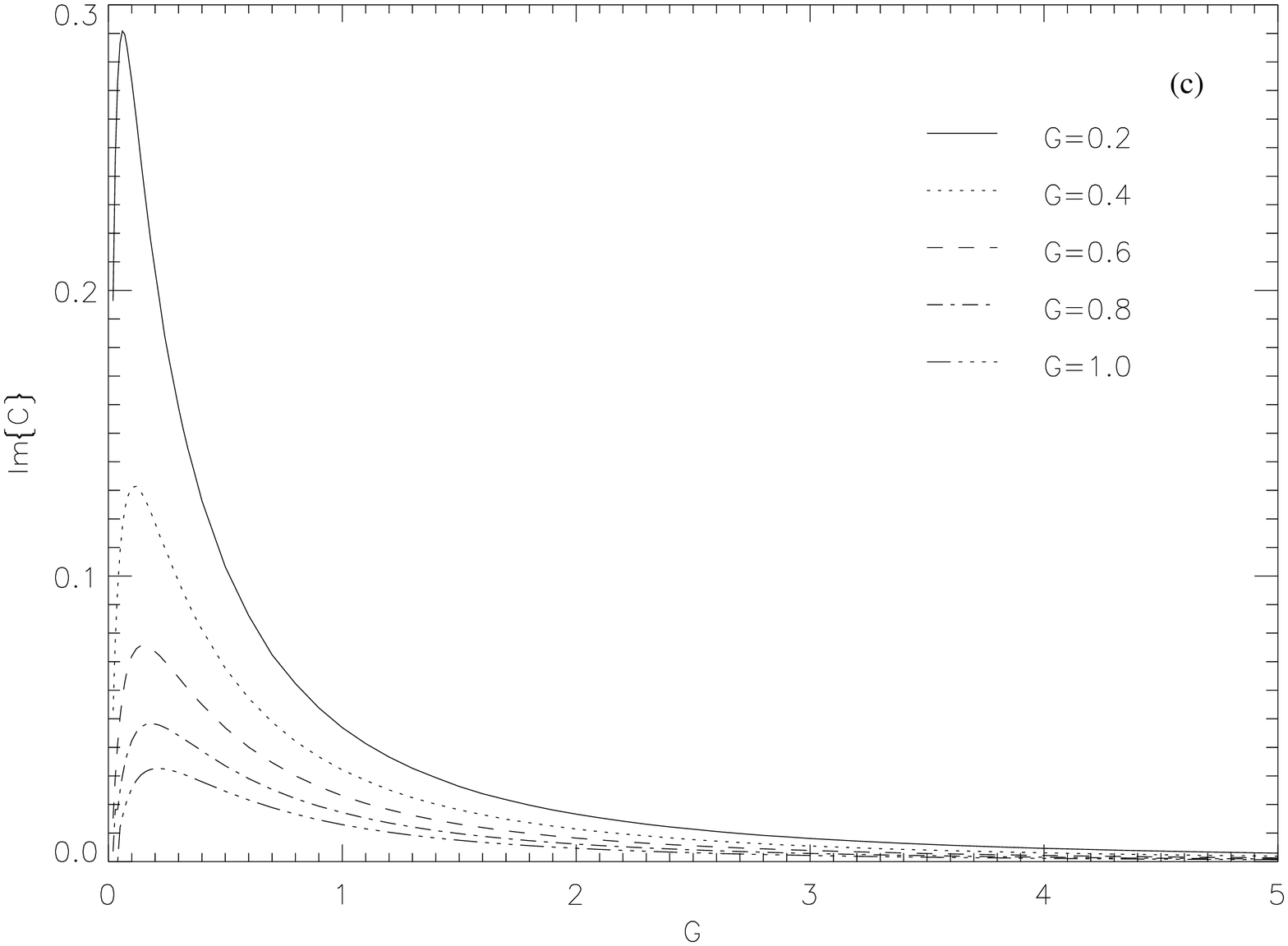}
\caption{Imaginary part of $C$ for a logarithmic wind profile
(a)$r=0.01$  (b)$r=0.1$  (c)$r=0.5$ .}
\label{fig2}
\end{figure}

\begin{figure}[htbp]
\epsfysize=2.5in
\epsfxsize=4.0in
\epsfbox{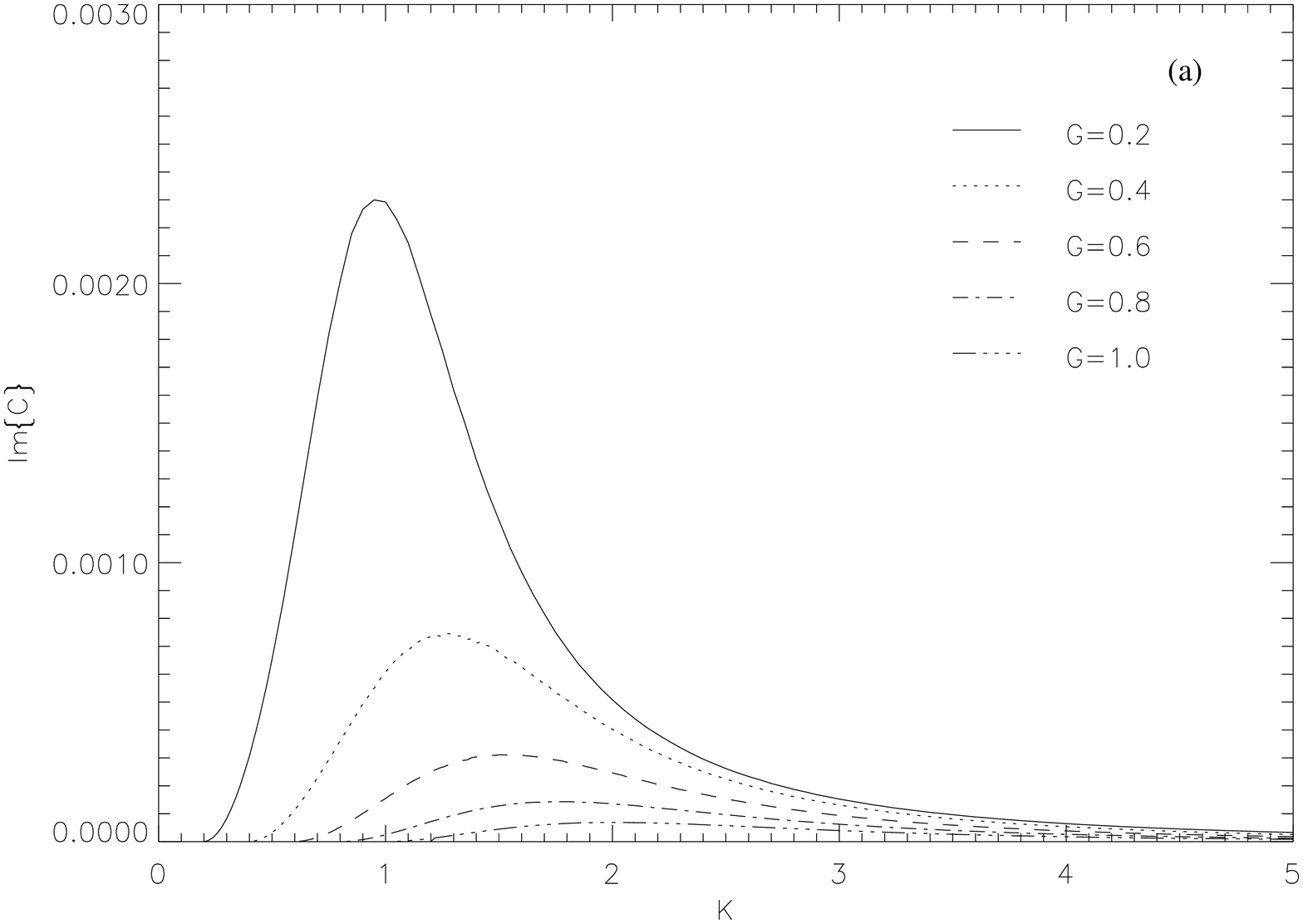}
\epsfysize=2.5in
\epsfxsize=4.0in
\epsfbox{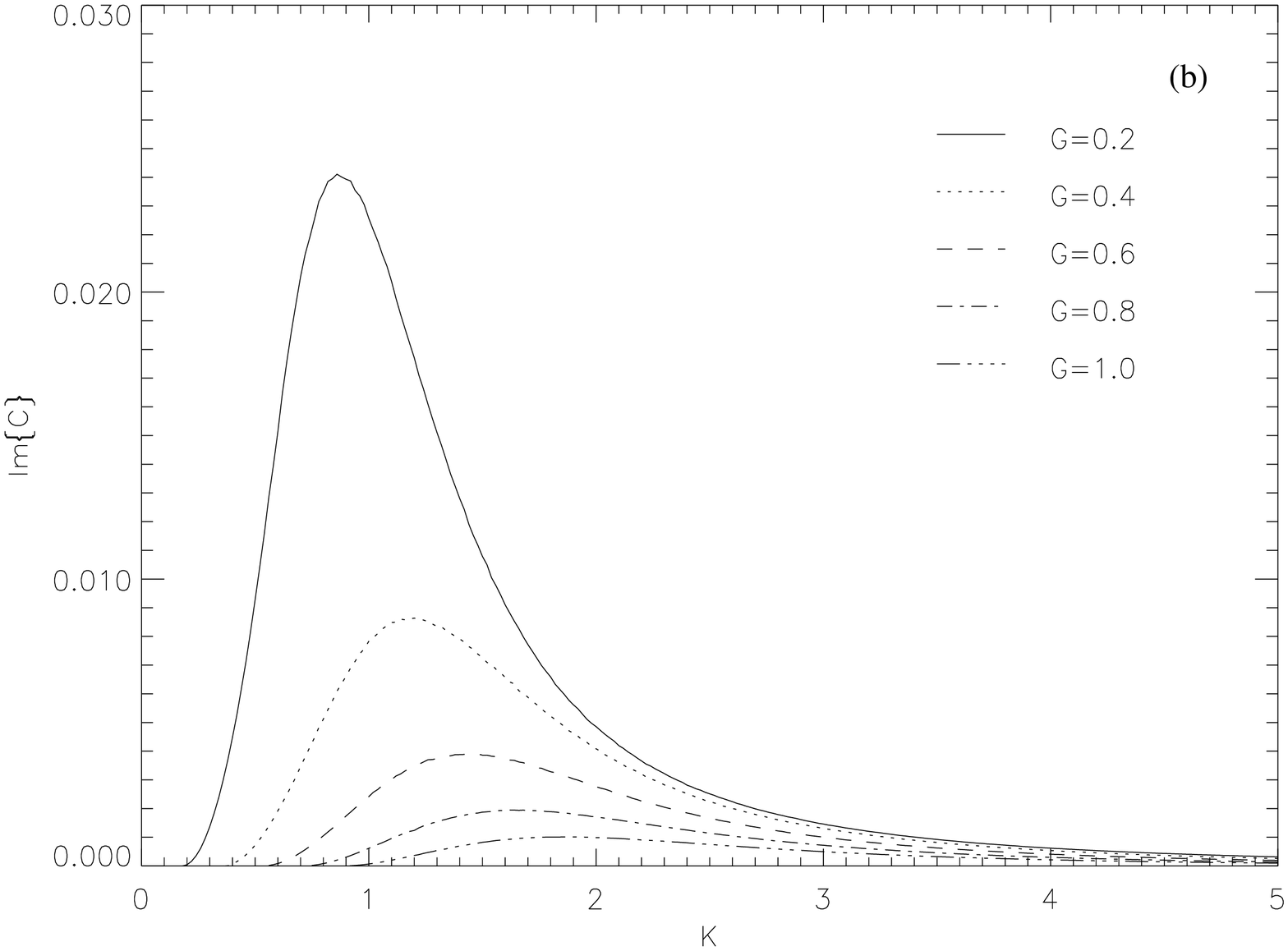}
\epsfysize=2.5in
\epsfxsize=4.0in
\epsfbox{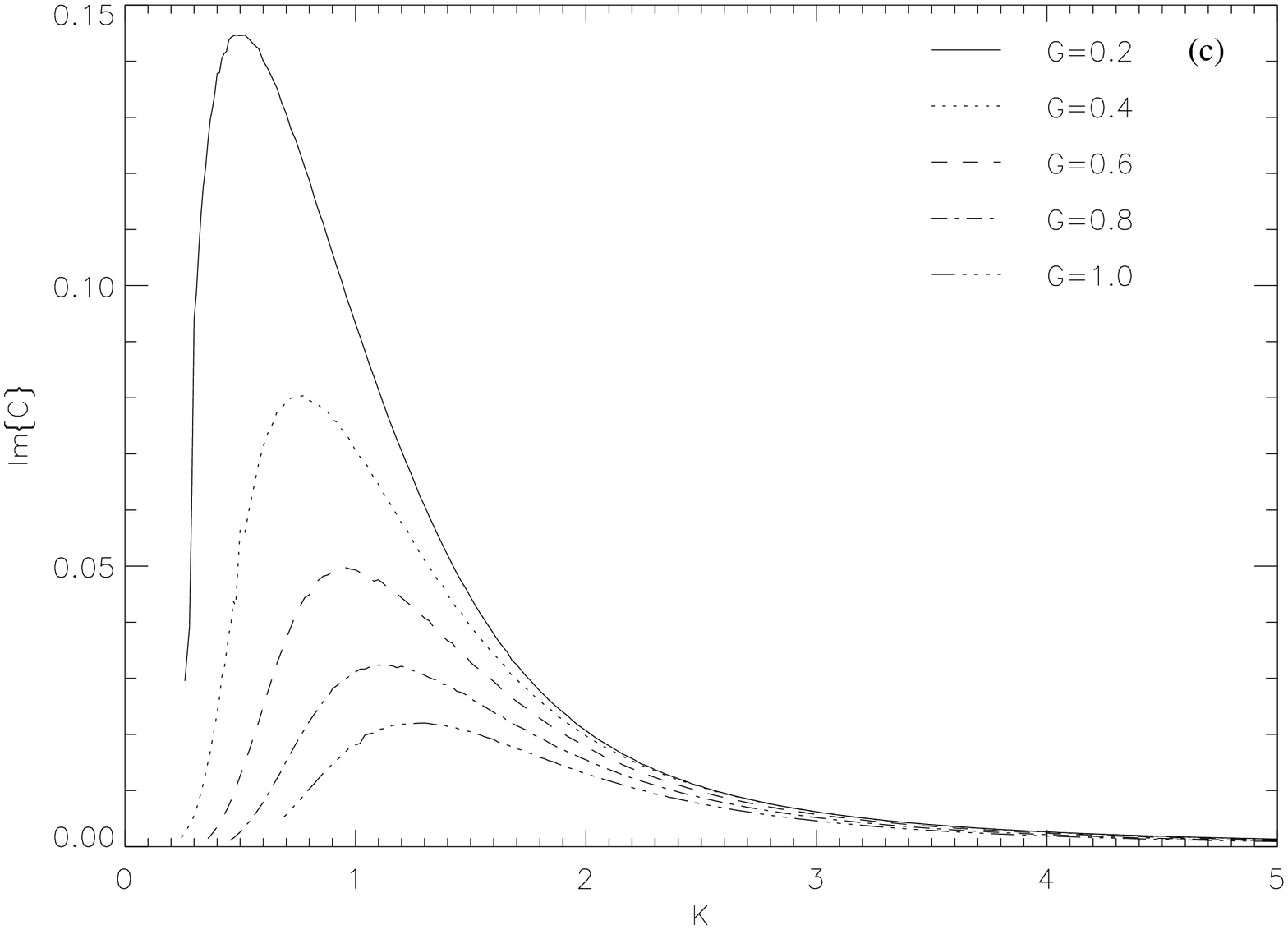}
\caption{Imaginary part of $C$ for a tanh wind profile
(a)$r=0.01$  (b)$r=0.1$  (c)$r=0.5$ .}
\label{fig3}
\end{figure}

For large density ratio, we solve the system of equations
\ref{linear_rescale_01}--\ref{linear_rescale_02} directly. We focus on
the instability properties of special interest, such as the maximum growth
rate, the wavelength of the fastest growing mode, and the dependence of
the the stability boundaries on the parameters of the model. First we
present results for cases where the wind has zero velocity at the
interface ($V_o=0$) in figures \ref{fig2}a-c and \ref{fig3}a-c.
  We solve equations \ref{linear_rescale_01}-
\ref{linear_rescale_02} numerically using a Newton-Ralphson method; the
results for both wind profiles logarithmic and tanh are presented together
 for
comparison. The plots sugest that for small enough $r$ the dependence on
$r$ is linear (e.g. the $r=0.001$ case is proportional to the $r=0.01$
case by exactly a factor of 10.0). For larger values of $r$ the
dependence is stronger than linear, and the smaller $K$ modes seem to
become more unstable.

\begin{figure}[htbp]
\epsfysize=2.5in
\epsfxsize=4.0in
\epsfbox{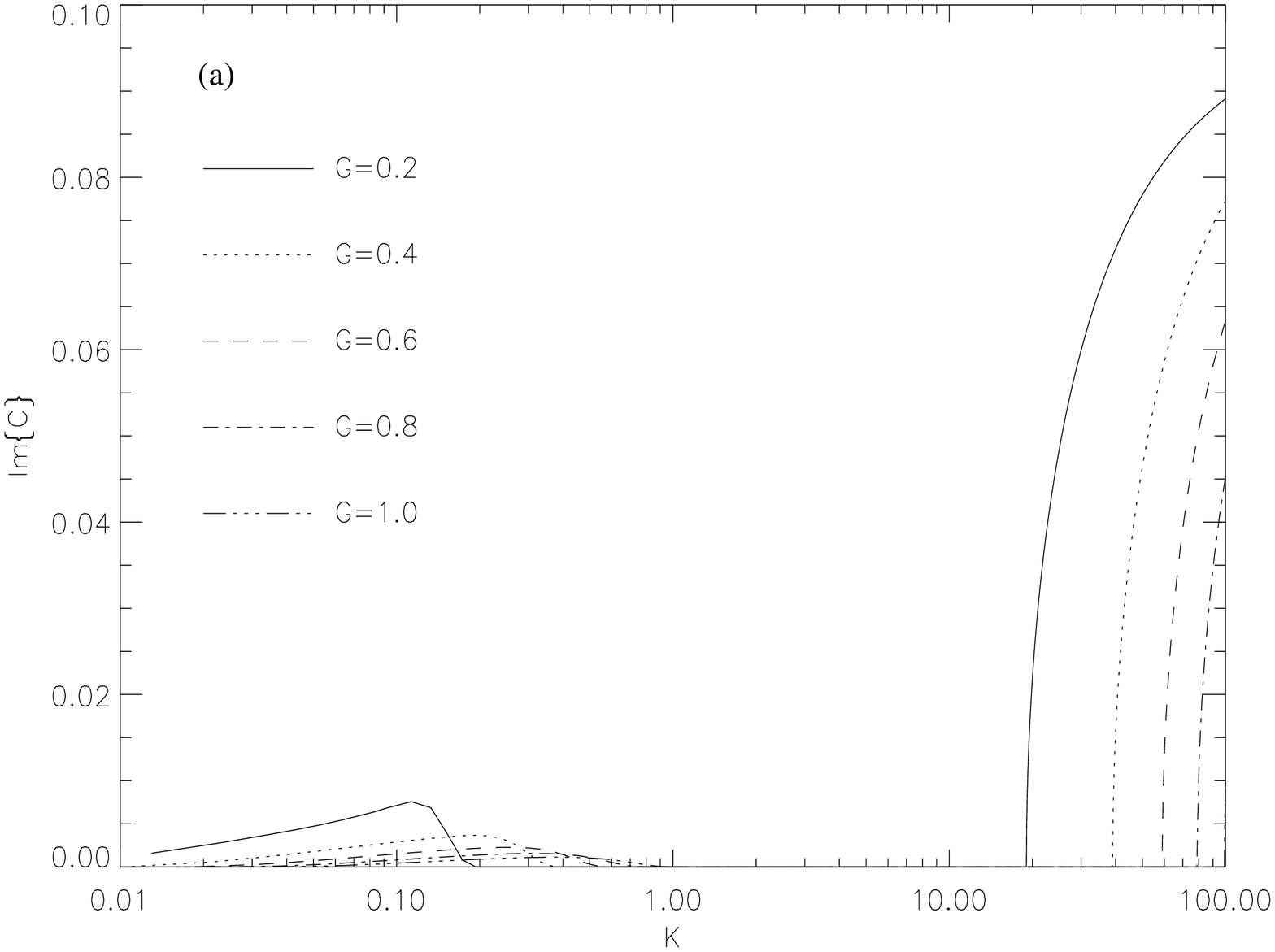}
\epsfysize=2.5in
\epsfxsize=4.0in
\epsfbox{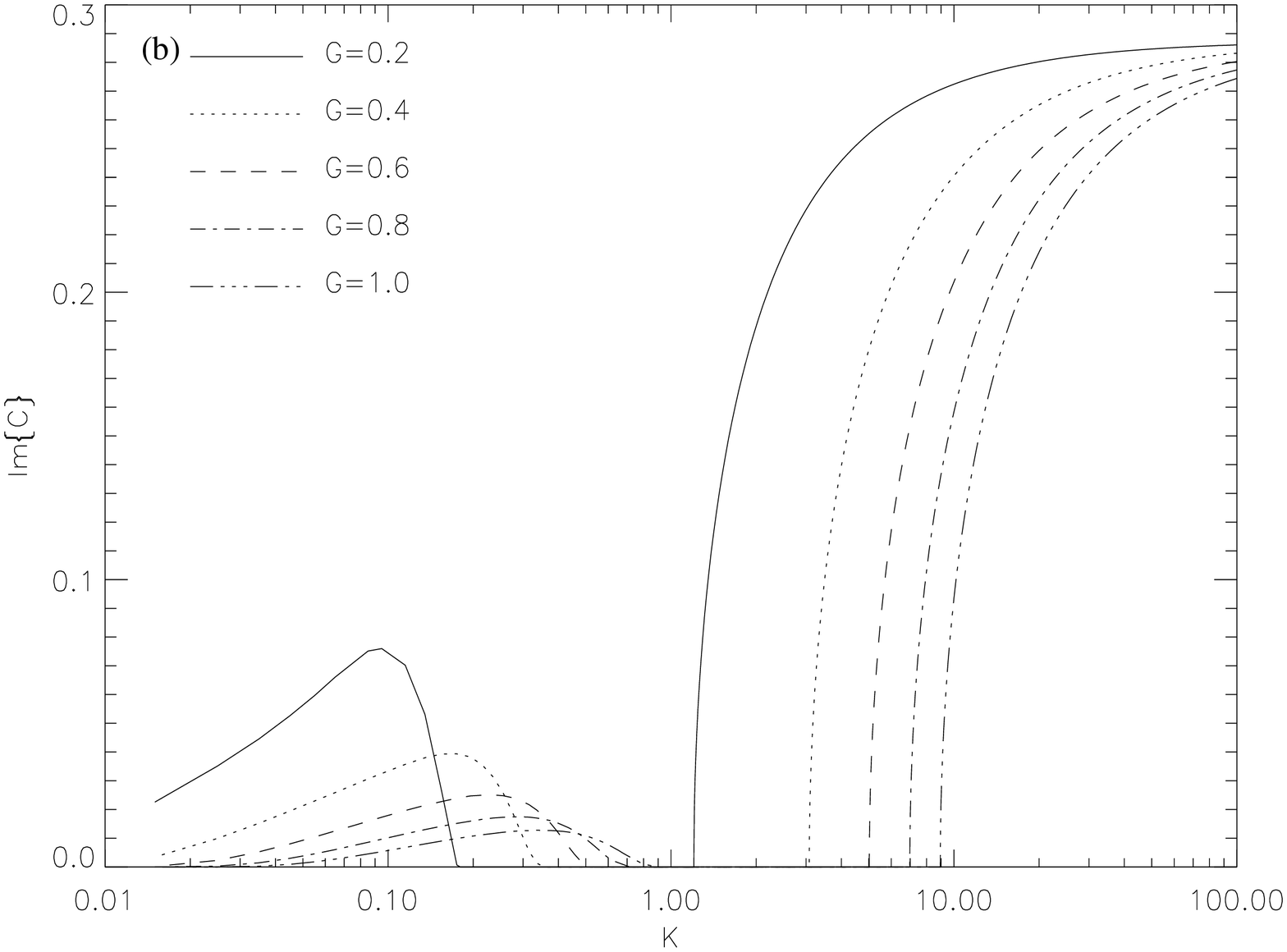}
\epsfysize=2.5in
\epsfxsize=4.0in
\epsfbox{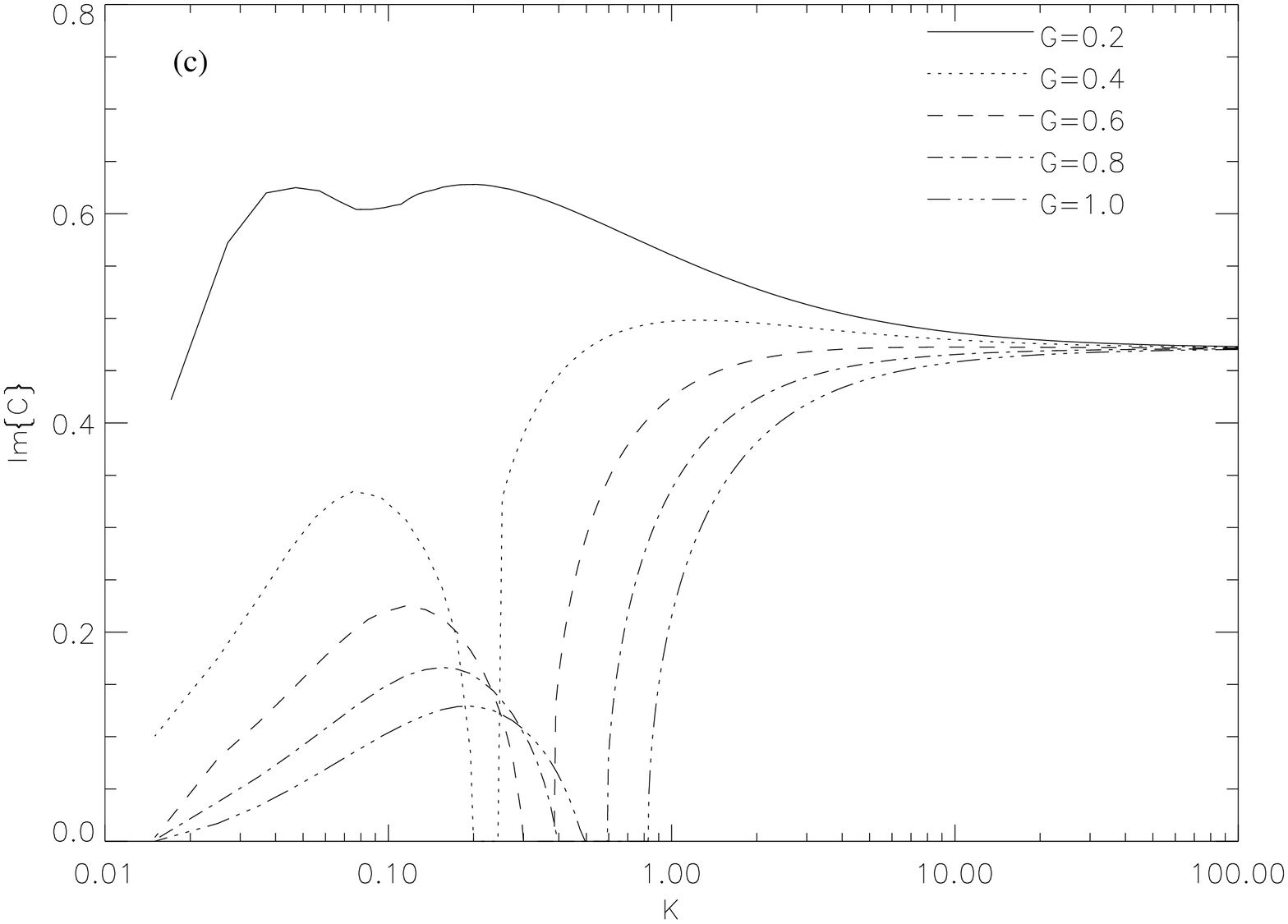}
\caption{Imaginary part of $C$ for a logarithmic wind profile and $V_o$=1.0
(a)$r=0.01$,  (b)$r=0.1$,  (c)$r=0.5$  .}
\label{fig4}
\end{figure}

\begin{figure}[htbp]
\epsfysize=2.5in
\epsfxsize=4.0in
\epsfbox{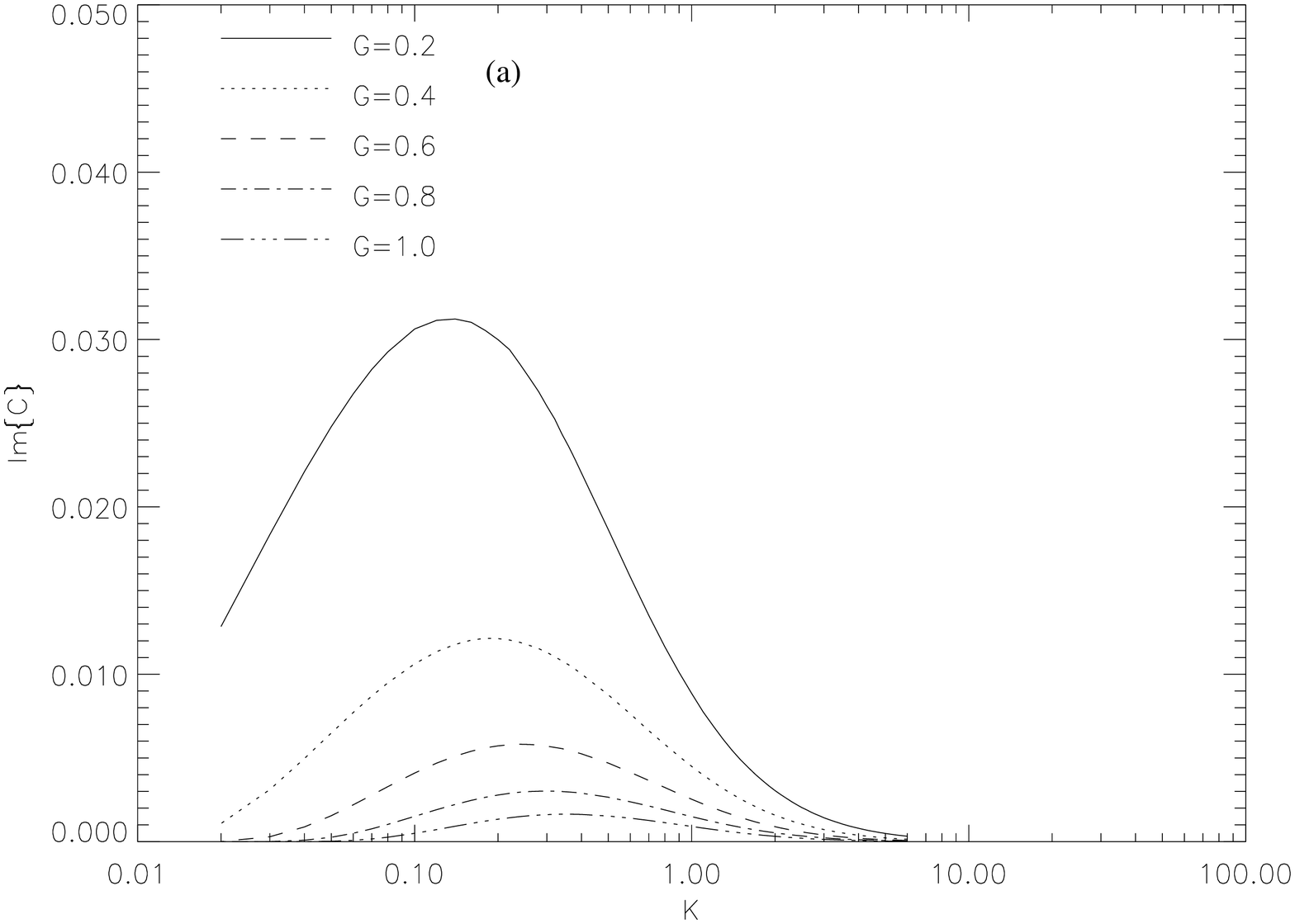}
\epsfysize=2.5in
\epsfxsize=4.0in
\epsfbox{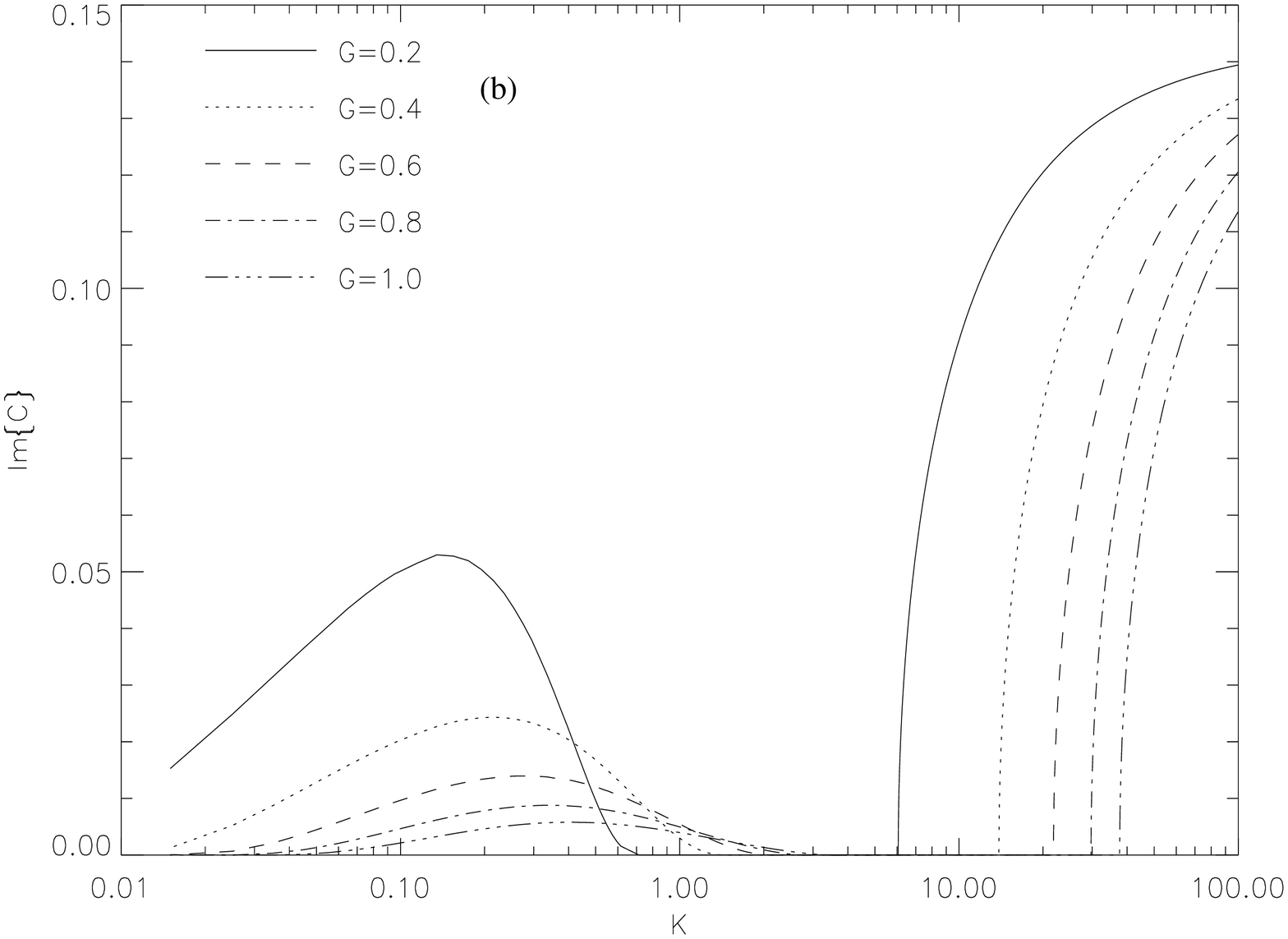}
\epsfysize=2.5in
\epsfxsize=4.0in
\epsfbox{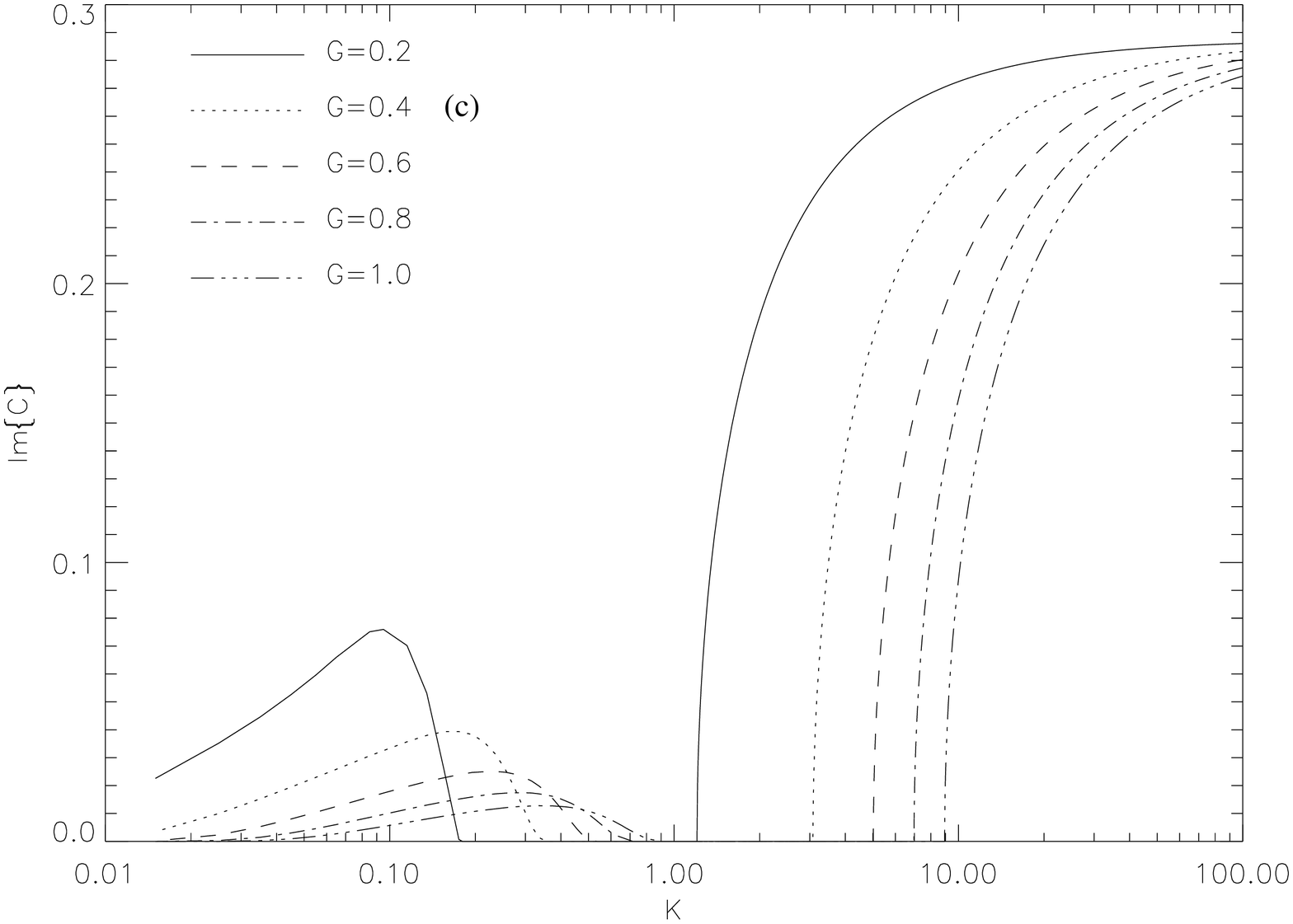}
\epsfysize=2.5in
\epsfxsize=4.0in
\epsfbox{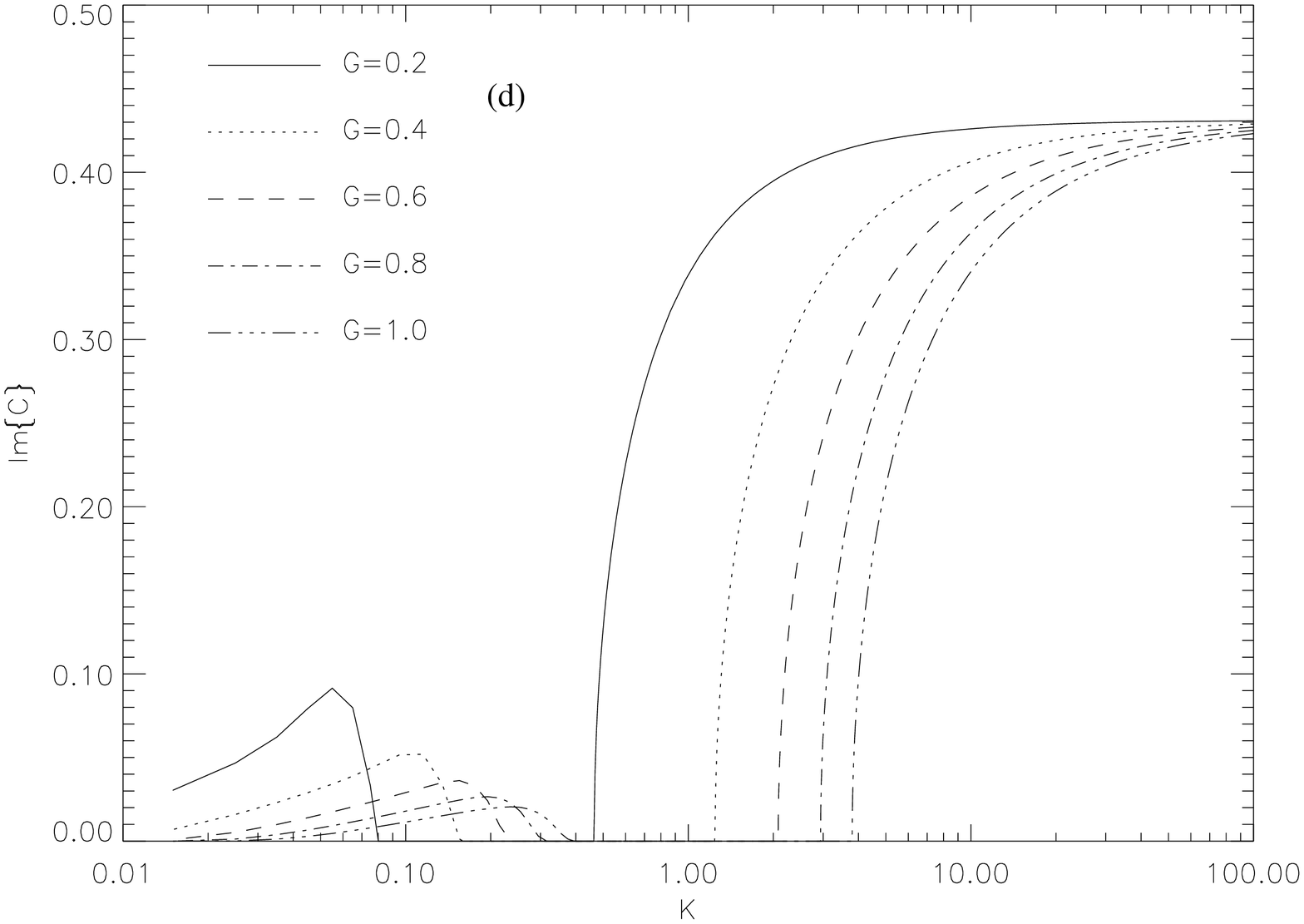}
\caption{Imaginary part of $C$ for a logarithmic wind profile with $r=0.1$
and (a) $V_o$=0.0,  (b)$V_o$=0.5,  (c)$V_o$=1.0,  (d) $V_o$=1.5  .}
\label{fig5}
\end{figure}

We have repeated these calculations for the case $V_o\neq 0$; the results
for the $\Im(C)$ are shown in figures \ref{fig4}a-c and \ref{fig5}a-d.
 In this case
we have to distinguish again the two different kinds of instabilities. The
distinguishing factor for the Kelvin-Helmholtz instability (most
prominent in the discontinuous wind profile) is that the growth rate is
positive only for wavenumbers larger than some lower bound. However, the
critical-layer instability, which owes its presence to the phase
change in the critical layer, has an upper bound in wavenumber for
instability. Thus in general there exists a band of wavenumbers for which
both modes are stable. The difference between small and large density
ratios is that the two instabilities are not necessarily separate,
and they do overlap for some parameters.
The criterion for a critical layer to exist in this case, $\sqrt{G(1-r)/K}
\ge V_o$,
provides a upper bound on $K$ for unstable CL modes. An exact solution
for the upper boundary is not available, but the asymptotic behavior of
the second boundary, for large $K$ and for small $V_o$ (cf., equation
\ref{expansion}) suggests that it takes the form of $K^{-1/2}$; thus, the
two stability boundaries are not expected to cross in the large
wavenumber limit. However, the two boundaries do meet for small $K$ and
large $V_o$, as it can be seen in the stability boundary plots
\ref{fig6}a-c.


\subsection{General features of the CL instability} \
\label{subsec:general_cl}
The main goal of this paper is to establish the relevance of the
critical-layer instability under various astrophysical or geophysical
conditions. Results from our linear analysis allow us to identify the
most unstable modes in different parameter regimes (and thus physical
situations). Furthermore, the maximum growth rates give us an estimated
time scale of the nonlinear evolution, and the length scale of the
fastest growing mode allow us to estimate the thickness of the mixing
layer as instability grows; this allows us to provide rough predictions
of the physical conditions for which more exact fully-nonlinear
calculations should be carried out to establish the mixing zone
properties more precisely. With this motivation in mind, we show here how
these two quantities behave as functions of the physical parameters
$G$, $r$ and $a$.

In figures \ref{fig7}a-b we have plotted the maximum growth rate
$\gamma\equiv\Im\{CK\}_{max}$ of the perturbation as a function of the
control parameter $G$ for the two wind profiles and for different values
of $r$. It is clear in all cases that there is an exponential dependence
on $G$ for $G\gtrsim 1$. 
This might be expected because increasing
gravity leads to an increase of the real part of $C$; therefore the 
imaginary part of $\phi$, that falls exponentially with the distance from 
the critical layer, will have an exponentially smaller component at the
 interface.
 Furthermore, as the
density ratio $r$ approaches unity, the dependence on gravity becomes
weaker. We plot the slopes of the curves from figures
\ref{fig7} as a function of $r$ in figure \ref{fig8}. The
dependence on $r$ is roughly linear (deviations from linearity will not be
important since $r$ only takes values in the range $0<r<1$). This allows
us to write an empirical scaling law law for the dependence of the growth
rate on the control parameters:
\begin{equation}
\label{empiric}
\gamma_{max} \equiv
KC_i\simeq \beta re^{-\alpha (1-r)G}
\end{equation}
For the logarithmic wind profile we found $\alpha \simeq 2.8$ and
$\beta \simeq 0.10$; while for the tanh wind profile we found
$\alpha \simeq 2.9$ and $\beta \simeq 0.18$.

In figures \ref{fig10}a,b we have plotted the wavenumber of the
most unstable mode (whose growth rate corresponds to equation
\ref{empiric}) as a function of $G$. We see that the wavelength of the
most unstable mode highly depends on the wind profile length scale. In
particular, for fixed density ratio $r$, $\lambda_{max} \sim a^{-1}$; the
dependence on gravity or on $r$ is weaker.


\section{Surface Tension}
\label{sec:tension}

For the sake of completeness, we have also examined the case in
which surface tension at the density interface is included\footnote{
We note that a magnetic field in the lower fluid, whose direction is
aligned with the flow,
would lead to an equivalent treatment (see for example
\cite{Chandra62} \S106).}.
 We again assume a wind shear
profile of the form $\ln(y+1)$ and tanh($y$). The only change in our set
of equations to solve is then an additional term in the boundary
condition, equation \ref{linear_rescale_02}. Hence

\begin{equation}
\label{linear_rescale_T}
KC^2-r\Big{[}(V_0-C)^2\phi_y-(V_0-C)V_y \Big{]}-G(1-r)-TK^2=0,
\end{equation}
where $T \equiv \sigma a/(\rho_2\su_1^2))$ and $\sigma$ is the surface
tension ($\sigma= B^2/(2\pi\mu K)$ for the case of the magnetic field
\cite{Chandra62}).
 We show the resulting solutions, namely the dispersion
relations, in figures \ref{fig11}a-c and \ref{fig12}a-c. 
 As expected tension decreases the growth rate and
becomes  important only in large wavenumbers.

An important result, which we have not previously seen derived, is that in
the small density ratio limit, the real part of $C$ (to zeroth order in
$r$) is $C_0=\sqrt{G/K+TK}$ which has its minimum value
$C_{min}=\sqrt{2}(GT)^{1/4}$ at $K=\sqrt{G/T}$. Thus, for the case of a
bounded wind profile (such as the tanh profile) there is a minimum value
of $U_1$, given by $C_{min}$, so that a critical layer can exist. We
remind the reader that a similar minimum velocity bound also exists for
the Kelvin-Helmholtz instability, and is given by

\[U\ge \sqrt{ \frac{2}{r}\sqrt{\frac{g\sigma}{\rho_2}}} \approx 650 \;
\mbox{ cm/sec} ~,
\]
where we have retained only terms of first order in $r$. For the CL case
we have instead

\[U\ge \sqrt{ 2\sqrt{\frac{g\sigma}{\rho_2}}} \approx 20 \; \mbox{
cm/sec} ~,
\]
which differs from the previous bound by a factor of $\sqrt{r}$. (The
numerical values shown here are derived for the case of air blowing
over water.) This illustrates the fact that for low wind conditions, the
CL instability dominates the KH instability for driving water surface
waves.


\section{Compressible case}
\label{sec:compress}

Finally we consider the compressible case. We will consider a compressible
fluid in the upper layer with sound speed $c_s(y)$ and an incompressible
 fluid below.
 The dimensionless equations
we have to solve now are:
\begin{equation}
\partial_y^2\phi-\left[ \kappa^2+G\frac{\tilde{\rho}}{\rho|_{y=0^+}}
\frac{K_s+K_g}{\tilde{V}_G^2}+
\frac{\partial_y^2\tilde{V}_G}{\tilde{V}_G}\right]\phi=0
\end{equation}
\begin{equation}
KC^2-r\Big{[}\tilde{V}_G^2|_{y=0}\partial_y\phi-\tilde{V}_G|_{y=0}
\partial_y \tilde{V}_G|_{y=0}
 \Big{]}-G(1-r)=0
\end{equation}
where

\[\kappa^2=K^2\left( 1-\frac{V_G^2}{C_s^2}\right), \; \;
K_s=\frac{\partial_y\rho}{a\rho}, \; \; K_g=G/C_s^2, \; \; C_s=c_s/U_1\]
and
\[\tilde{V}_G=\frac{KV_G\sqrt{\tilde{\rho}/\rho|_{y=0^+}}}{\kappa}
\; \; \mbox{with} \; \; r=\rho|_{y=0^+}/\rho|_{y=0^-}. \]

We will assume for simplicity an adiabatic atmosphere,
\begin{equation}
\rho=\rho|_{y=0^+} \left( 1-(\gamma-1)\frac{\rho_o}{\gamma P_o}gy \right)^
{\frac{1}{1-\gamma}}
\end{equation}

\begin{equation}
P=P|_{y=0} \left(\frac{\rho}{\rho_o}\right)^\gamma.
\end{equation}
This assumption, which is commonly used in the atmospherical sciences to
simplify the physics involved, has the advantage that $K_g+K_s=0$, so our
equation becomes by one order less singular, and therefore becomes
easier to solve.

We will not deal here with supersonic flows since in most astrophysical
realms in which interfacial wave generation plays an important role
(viz., on white dwarf surfaces) the relevant flows are thought to be
subsonic; for this reason we will consider only the tanh wind profile. We
shall also deal with small values of $G$, so that the pressure scale
height is large and the breakdown of the adiabatic assumption at values
of $y\sim K_s^{-1}$ will not affect us either.

The dispersion relation for different values of $C_s$ and for a tanh wind
profile is given in figures \ref{fig13}a,b. Compressibility, as
it can be seen from the figures, decreases the growth rate. This is an
expected result, since our system has now more degrees of freedom (e.g.,
now the perturbation stores thermal energy as well). We conclude, however,
that the deviation from the incompressible case is not very large, even
for relatively strong (but still subsonic) winds.

\section{Discussion and Conclusion}

In this paper we have explored the linear instability properties of wind
shear layers in the presence of gravitational stratification. Our
principal aim was to explore the full parameter space of the solutions,
defined by the four parameters $K$ (the perturbation wavenumber), $G$
(related to the Richardson number, and measuring the relative energy
contributions of the gravitational stratification and the wind), $V_o$
(the velocity discontinuity at the density interface), and $r$ (the
 density ratio).

 We have distinguished between the two different kind of modes
(Kelvin-Helmholtz modes and Critical Layer modes) existing in our model 
and constructed stability boundaries for those, as well as the dependence of
 these 
boundaries on the given parameters. As we discuss later, the non-linear 
development of the instability (and therefore mixing) will crucially 
depend on the kind of modes that become unstable; therefore an
 investigation of
the stability boundaries is necessary before the study of the  
nonlinear regime.
An important result also derived from our analysis, allowing us to
make predictions on the importance of the instability and on the nonlinear  
development, is the scaling of the 
growth rate with the parameters $G$ and $r$ 
in \S \ref{subsec:general_cl}; our results show that for 
$(1-r)G\gg 1$ strong mixing is not expected.
This result has an interesting implication: although, as expected, strong
gravity (e.g. $G$) inhibits mixing, one might still experience strong
instability in the large $G$ case if the density ratio of the interface is
close to but not equal to unity.
Finally we investigated the effects of surface tension and compressibility.
With the inclusion of surface tension, we obtained a lower 
bound on $U_{max}$ for the instability to exist.
We also found that for subsonic winds the instability growth rate
weakly depends on the Mach number.

As we have shown, there are significant differences between the CL and
the KH modes, both in the parameter ranges in which the instability can occur
(e.g., the stability boundaries) and in the magnitude of the growth rate;
these differences can be expected to result in different nonlinear
evolution of the underlying physical system. For example, it is
well-known that CL instability in the air over water case is responsible
for generating waves for winds of magnitude below the threshold for which
Kelvin-Helmholtz instability exists \cite{Miles57}.

An important aspect not discussed as yet is the case in which the spatial
density variation is smooth instead of discontinuous. In our simplified
model of a sharp interface, the distinction between CL modes and KH modes
emerges naturally from our analysis, simply based on the existence or
absence of a critical layer. In the more realistic (astro)physical case,
however, sharp velocity and density gradients do not exist. For this
reason, we need to generalize our definitions for the two kinds of modes.
We proceed by considering the physical mechanisms involved in the
instabilities:
In the KH case, as mentioned above, the pressure
perturbations are in phase with the gravity wave amplitude; and the wave
becomes unstable when pressure overcomes the restoring force (e.g.,
gravity). An immediate consequence of this is that when the restoring
force is overcome, it no longer plays a role in the wave propagation,
and therefore the real part of $c$ is independent of the restoring force,
i.e., independent of $g$. This argument can be confirmed by examining
the results for the step-function wind profile, where we see precisely the
predicted behavior.

In the CL case, it is instead the out-of-phase pressure component that
drives the wave unstable; in this case, the destabilizing pressure
force does not strongly modify the restoring force (here again gravity).
Hence the real part of $c$ is only weakly modified when the mode
becomes unstable, and therefore the wave continues to propagate with its
``natural'' speed while going unstable. These properties of wave
destabilization, which affect the dependence of the real part of
$c$ on the restoring force, can therefore be used to distinguish between
the CL and KH modes. Thus, in the more general case, we shall refer to
the modes that become unstable due to an in-phase pressure component as
KH modes; their propagation speed is independent of (or at most weakly
dependent on) the restoring force. In contrast, we shall refer to the
modes that become unstable due to an out-of-phase pressure component
as the ``resonant modes" (since the name ``critical layer" is not
appropriate for the general case); their propagation speed depends
strongly on the restoring force.\footnote{The words `weakly' and
`strongly' are used here because it is expected that there is a smooth
transition from the one case to the other as we change our physical
parameters.}

By extrapolating our results to the smooth density variation case we
conclude that KH modes are likely to appear in cases in which the 
inflection point of the wind profile, or the region in space in which 
$U$ changes, is at the same height as the region which the density changes.
We note that the KH instability, as defined in \S \ref{subsec:KH_vs_CL}, 
is a limiting case 
of such wind profiles. In contrast, resonant modes are more likely to
 appear when
the regions of velocity and density change are well separated,
 where the coupling
between an existing gravity wave and a critical layer above can lead to a
`resonant' behavior as described above.
This expectation is
supported even further by the observation that in the case of a smoothly 
stratified fluid the stratification term becomes dominant in the critical 
layer and the phase change is no longer $i\pi$ but depends on the Richardson 
number. 

We also note that, in the past literature on shear flow instability,
much attention is focused on the KH instability according to our definition. 
For example, models with $U \sim \tanh(y)$ and $\partial_y \ln(\rho) \sim
\tanh(y)$ or $\rho \sim \exp (-y)$ have been studied in the linear, weakly
nonlinear, and fully nonlinear regimes \cite{Drazin81,nlkh3,nlkh2,nlkh1},
 but they all fall into the KH
category (as we defined it); complete study of resonant modes, though, has not
been fully investigated. This is a gap we seek to close in our further
work.

\begin{figure}[htbp]
\epsfysize=2.5in
\epsfxsize=4.0in
\epsfbox{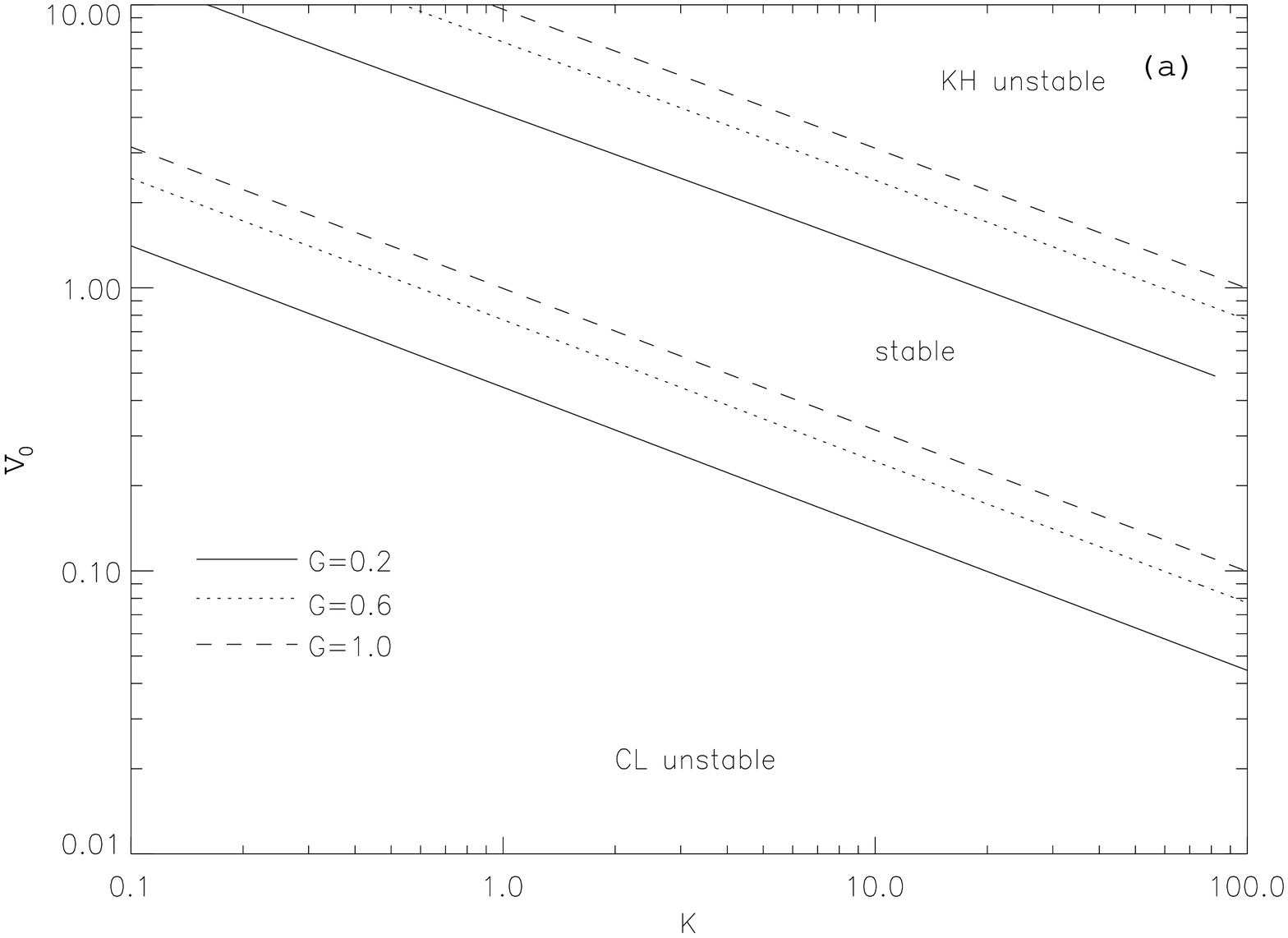}
\epsfysize=2.5in
\epsfxsize=4.0in
\epsfbox{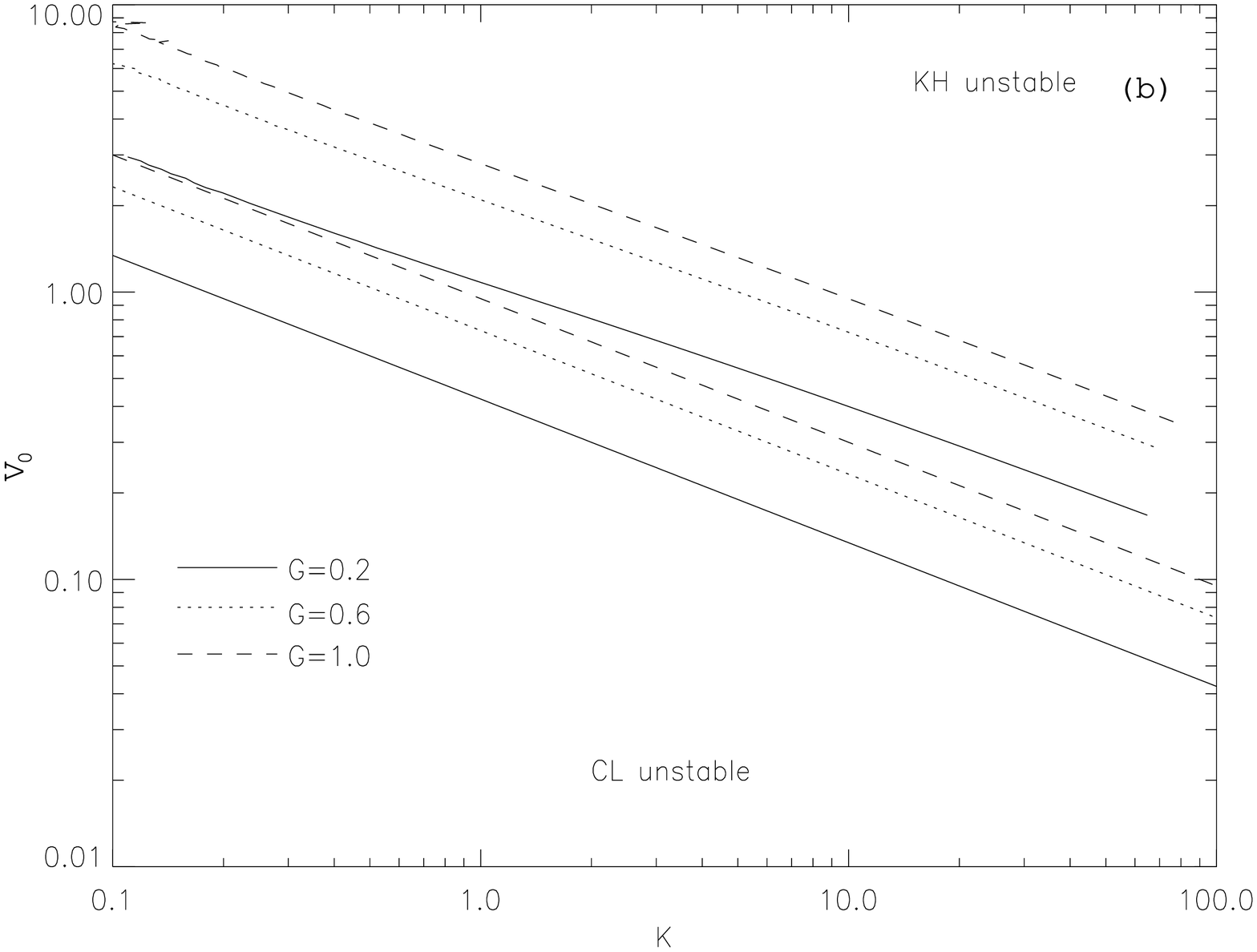}
\epsfysize=2.5in
\epsfxsize=4.0in
\epsfbox{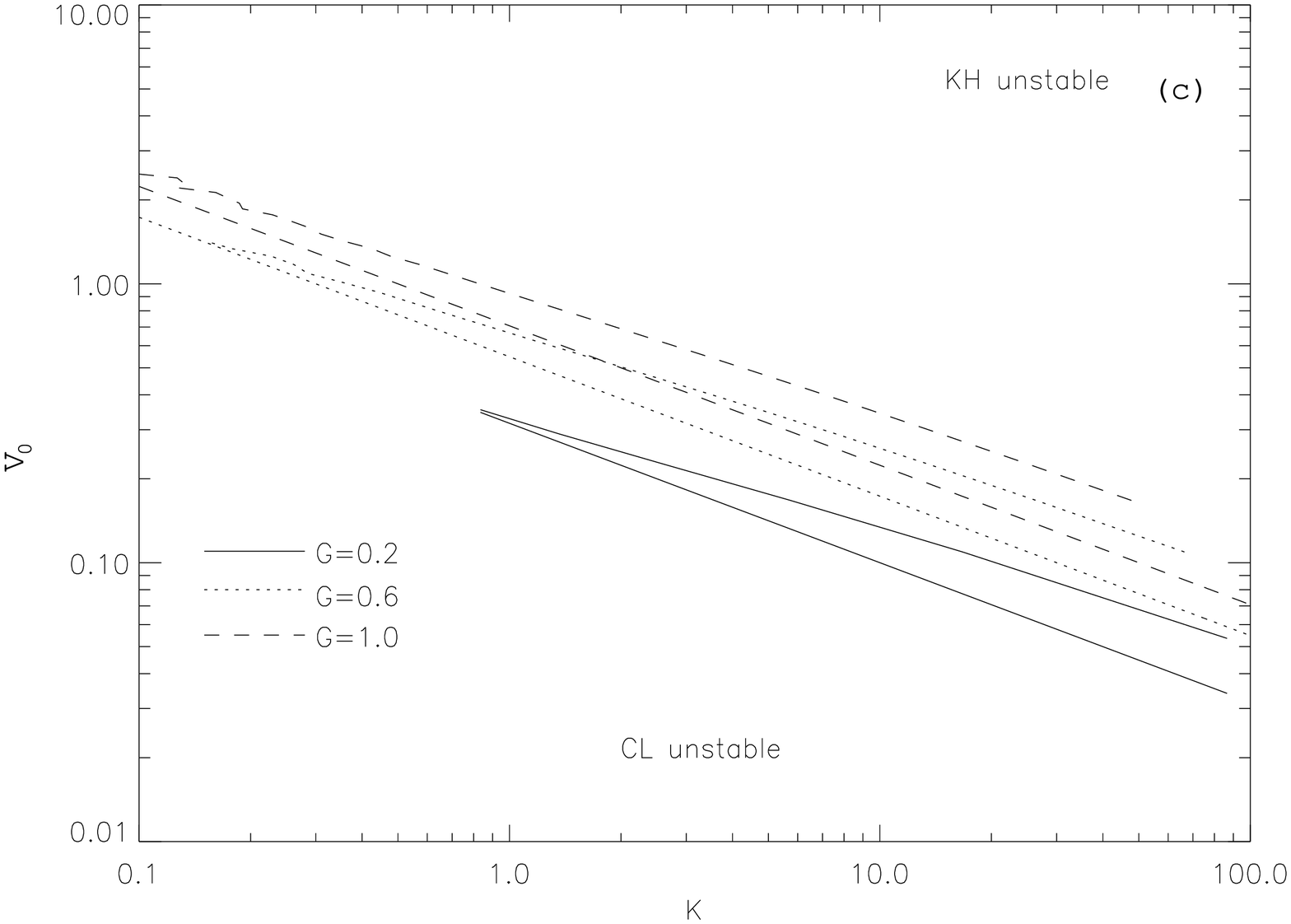}
\caption{Stability boundaries for (a)$r=0.01$, (b)$r=0.1$, (c)$r=0.5$ .}
\label{fig6}
\end{figure}

\begin{figure}[htbp]
\epsfysize=2.5in
\epsfxsize=4.0in
\epsfbox{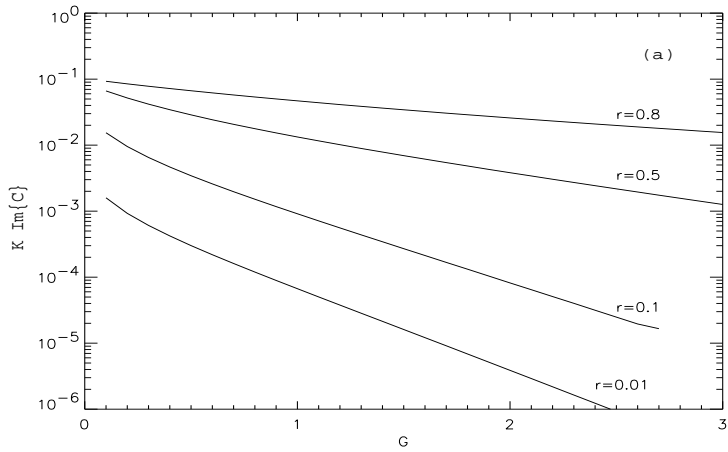}
\epsfysize=2.5in
\epsfxsize=4.0in
\epsfbox{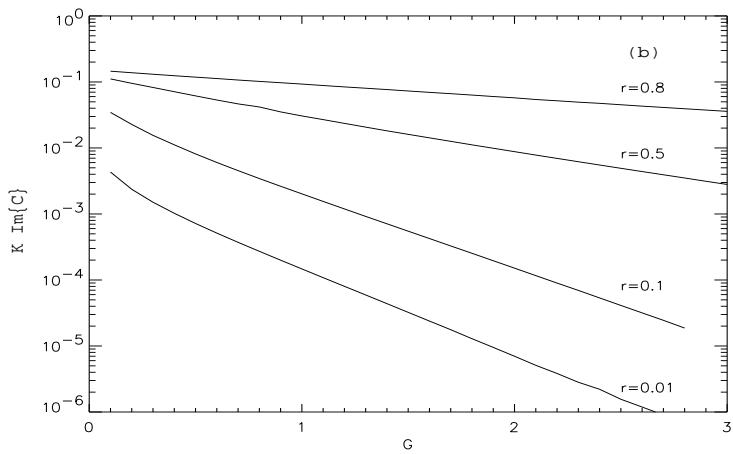}
\caption{Growth rate of the fastest growing mode as a function of $G$
(a)logarithmic wind profile,  (b) tanh wind profile .}
\label{fig7}
\end{figure}

\begin{figure}[htbp]
\epsfysize=2.5in
\epsfxsize=4.0in
\epsfbox{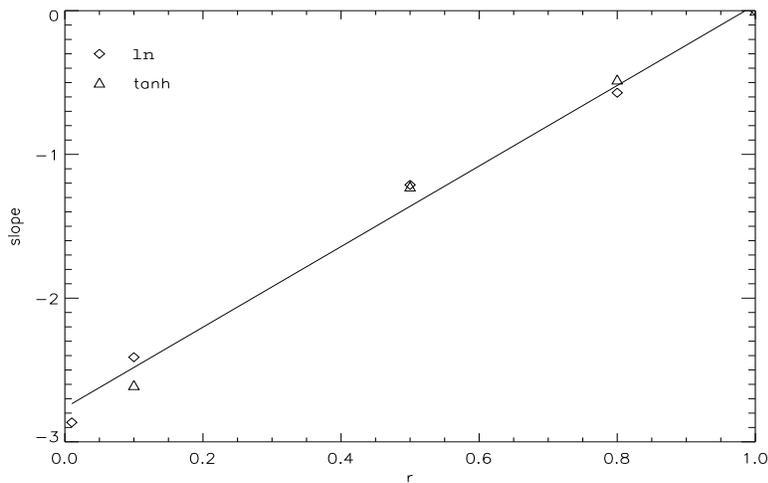}
\caption{Slopes of the previous graph as a function of $r$
(e.g. the dependence of the exponent in eq. \ref{empiric} on $r$).}
\label{fig8}
\end{figure}

From our results and the ranges of our physical parameters in Table
\ref{table1} we can estimate the growth rates of the wind driven waves
as well as their typical wave length. For the astrophysical problem we are
interested in, we conclude that the growth rate can be as large as
$10^{-2}$s$^{-1}$ with typical wave lengths of the order of $a^{-1}$;
these results were obtained using the empirical formulae \ref{empiric}
and \ref{bound1}. For our motivating astrophysical application, i.e., the
nova mixing problem, the results shown in \ref{fig6} are especially
important. First, we note that the interface between the stellar surface
(at the typical density $\rho_2=3800 gcm^{-3}$) and the accreted
envelope (at the typical density $\rho_1=400 gcm^{-3}$) is a material
(gaseous) boundary at which one would not expect any free slip. Thus, we
would expect $V_o \equiv U_o/U_1$ (in, for example \ref{windpr}) to be
very small, and essentially zero. Consider then Panel(b) of \ref{fig6} 
(for which $r$ takes on the astrophysical relevant value): we see that for
small $V_o$, the interface instability is completely dominated by the
resonantly-driven modes; the classical KH instability only appears at very
large wave numbers, e.g. very short wavelengths and therefore is unlikely
to matter in the nova mixing problem. To go beyond this will require
further investigation of the nonlinear evolution of the CL instability and
is currently under investigation; more information on the
astrophysical model, is provided in \cite{Rosner00}. We discuss some of
the issues relevant to the nonlinear regime next.

Finally, we comment on the possible nonlinear development of the two
types of modes we have studied. The nonlinear evolution of a KH-mode is
in fact well-studied in the literature \cite{nlkh3,nlkh2,nlkh1}
 and is known  to lead to a
mixed region of width roughly equal to the wavelength of the mode;
indeed, mixing proceeds in this case until (roughly speaking) the
Richardson criterion holds throughout the flow. In the case of CL modes,
the nonlinear evolution is affected by the fact that a new length scale
enters the problem, namely the height of the critical layer, $y_c$, which
can be substantially larger than the mode wavelength. Thus, one might
expect that mixing proceeds until heights of order $y_c$ are reached by
the mixing layer, and therefore we expect more extensive mixing. Clearly,
the next steps in this study involve investigation of the weakly
nonlinear regime (to examine supercriticality and possible saturation of
the modes), as well as the fully nonlinear regime (through numerical
simulations). A particularly interesting question is to what extent the
expected wave breaking of the CL modes (cf. \cite{Rosner00}) can lead to
enhanced mixing at the shear/density interface.

\acknowledgments

We thank N. Balmforth, T. Dupont, R. Pierrehumbert and
J. Truran for helpful discussions.
AA and RR acknowledge the support of the DOE-funded ASCI/Alliances Center for
Astrophysical Thermonuclear Flashes located at the University of Chicago.
YY acknowledges support from NASA and Northwestern University.

\appendix
\section{Extension of Howard's semicircle theorem}
We begin from

\begin{equation}
\phi_{yy} - \Big{(}{K}^2 +\frac{V_{yy}}{V-C}
\Big{)}\phi=0,\quad \phi|_{y=0}=1,\quad
\phi|_{y=\infty}=0
\end{equation}
and
\begin{equation}
KC^2\phi-r\Big{[}(V_0-C)^2\phi_y-(V_0-C)V_y\phi \Big{]}-\tilde{G}\phi=0.
\end{equation}
where $\tilde{G}$ is the restoring force ($\tilde{G}=G(1-r)$ for the simplest
 case),
$0<\tilde{G}$, $0<K$, and we
assume $C_i\not= 0$. Let $V_G=V-C$, and let $\psi=\phi/V_G$ and
$D\equiv\partial_y$. Note that
\[
V^2_GD\phi-V_G\phi DV_G=V^3_GD\psi ~.
\]
The boundary condition can then be written as
\begin{equation}
rV^3_GD\psi=KC^2-\tilde{G} ~.
\end{equation}
 From (A1) we obtain
\begin{equation}
V_GD^2\psi+2DV_GD\psi-V_GK^2\psi=0 ~;
\end{equation}
multiplying the last relation with $V_G\psi^*$ and integrating, we obtain:
\[
\psi^*V^2_GD^2\psi+\psi^*DV^2_GD\psi-V^2_GK^2|\psi|^2 =0 ~,
\]
so that
\[
D[\psi V^2_GD\psi]-V^2_G|D\psi|^2-V^2_GK^2|\psi|^2 =0
\]
and
\[
-[\psi^*V^2_GD\psi]_0-\int_0^\infty V^2_G\big{[}|D\psi|+K^2|\psi|^2
\big{]}dy =0 ~;
\]
using the normalization condition, and denoting by
$Q(y)=\big{[}|D\psi|+K^2|\psi|^2\big{]} \ge 0$, we then have

\[ \frac{1}{V_G^*}V^2_G\frac{KC^2-\tilde{G}}{V_G^3}=
-r\int^\infty_0(V_G)^2Qdy ~,
\]
so that
\begin{equation}
\frac{KC^2-\tilde{G}}{|V_G|^2}=-r\int^\infty_0(V-C)^2Qdy ~.
\end{equation}
Taking the imaginary part, we obtain
\[
\frac{K(2C_rC_i)}{|V_G|^2}=r\int^\infty_02C_i(V-C_r)Qdy
\]
\begin{equation}
\frac{KC_r}{|V_G|^2}=r\int^\infty_0(V-C_r)Qdy ~.
\end{equation}
Therefore,
\begin{equation}
0<C_r<V_{max}
\end{equation}
i.e., a wind cannot generate waves traveling faster than its maximum
speed. Now, taking the real part, we obtain

\begin{equation}
\frac{K(C_r^2-C_i^2)-\tilde{G}}{|V_G|^2}=
-r\int^\infty_0 \big{[}V^2-2VC_r+C_r^2-C_i^2 \big{]}Qdy ~,
\end{equation}
or

\[\frac{K(C_r^2-C_i^2)-\tilde{G}}{|V_G|^2}=
-r\left[
\int^\infty_0V^2Qdy-2C_r\int^\infty_0VQdy
+(C_r^2-C_i^2)\int^\infty_0Qdy \right] ~,
\]
or
\[\frac{K(C_r^2-C_i^2)-\tilde{G}}{|V_G|^2}=
-r\left[\int^\infty_0V^2Qdy-2C_r
\left\{\frac{KC_r}{r|V_G|^2}+C_r\int^\infty_0Qdy \right\}
+(C_r^2-C_i^2)\int^\infty_0Qdy \right] ~,
\]
so that
\begin{equation}
0<\frac{K|C|^2+\tilde{G}}{|V_G|^2}=
r\int^\infty_0\big{[}V^2-|C|^2\big{]}Qdy
\end{equation}
and
\begin{equation}
C_r^2+C_i^2<V^2_{max} ~,
\end{equation}
which is the sought-for result.


\section{lower bound on the CL-unstable modes}

Consider the wind profile $V=1-e^{-y}$; then the equations
\ref{linear_rescale_01} and \ref{linear_rescale_02} become 

\begin{equation}
\label{bnd1}
\phi_{yy}-\left( K^2-\frac{e^{-x}}{1-e^{-x}-C} \right)\phi=0
\end{equation}
and
\begin{equation}
\label{bnd2}
C^2-r [ C^2 \phi_y +C ] -G(1-r)=0   .
\end{equation}

\noindent
We are interested in the value of $K$ for which our system becomes marginaly
 unstable.
From the extension of Howard's semicircle theorem to our case we know that
for $C_r$ greater than $V_{max}$ the system is stable, so the instability
is expected to start when $C=V_{max}=1$. Using this value for $C$ we
obtain from \ref{bnd1}

\begin{equation}
\phi_{yy}-\left( K^2+1 \right)\phi =0 \; ;
\end{equation}
therefore
$\phi=e^{-y\sqrt{K^2+1}}$ and from \ref{bnd2} we then have
\begin{equation}
K-r [-\sqrt{K^2+1}+1]-G(1-r)=0 \;,
\end{equation}
which by solving gives us

\begin{equation}
K_{min}=\frac{G(1-r)+r-r\sqrt{(G(1-r)+r)^2+(1-r^2)}}{1-r^2} \; .
\end{equation}
Numerical integration confirms this result.


\section{KH-modes in the small $\rho_1/\rho_2$ limit}

We begin with the Rayleigh equation \ref{linear_rescale_01}
for large $K$
\[ \phi_{yy}-\left( K^2+\frac{V_{yy}}{V-C} \right) \phi=0 ~.
\]
Set $z=Ky$ and $ \epsilon = 1/K $; we then have

\[
\phi_{zz}-\left( 1+\epsilon^2\frac{V_{zz}(\epsilon z)}
{V(\epsilon z)-C}\right) \phi=0
\]
or

\[
\phi_{zz}-\left(1+\epsilon^2 F(\epsilon z)\right)=0 ~,
\]
where $F(x)=V_{zz}(x)/(V(x)-C)$. Expanding $\phi$ in powers of
$\epsilon^2$,
\[
\phi=\phi_0+\epsilon^2 \phi_1+\epsilon^4\phi_2+... \; ,
\]
we obtain, to first order,
\[
\phi_0=e^{-z} ~;
\]
at the next order, we have
\[
\phi_{1zz}-\phi_1=F(\epsilon z)\phi_0 ~,
\]
which has the solution
\[
\phi_1=\int G(|z-x|)F(\epsilon x)\phi_0(x) dx+Ae^{-z}~,
\]
where $G=-\frac{1}{2} e^{-|x-z|}$ is the Greens function and $A$ is a
constant to be chosen so that $\phi_1$ will satisfy the boundary
condition at zero:

\[\phi_1(z)=-\frac{1}{2}\int_0^z e^{-(z-x)} e^{-x} F(\epsilon x) dx -
\frac{1}{2}\int_z^{\infty} e^{-(x-z)} e^{-x} F(\epsilon x) dx
+Ae^{-z} \]

\[=-\frac{1}{2}e^{-z}\int_0^zF(\epsilon x) dx -
\frac{1}{2}e^z\int^{\infty}_ze^{-2x}F(\epsilon x)dx+
\left( \frac{1}{2}\int^{\infty}_0e^{-2x}F(\epsilon x)dx \right)e^{-z} \]

\[=-\frac{1}{2}e^{-z}\Bigg{[}
\int_0^zF(\epsilon x) dx+
\int^{\infty}_0e^{-2w}F(\epsilon(w+z))dw
-\int^{\infty}_0e^{-2x}F(\epsilon x)dx \Bigg{]} \qquad (w=x-z) ~.
\]

\noindent
We are interested in the first derivative of $\phi$ at zero, so we can
obtain
\[\frac{d\phi_1}{dz}\Big{|}_{z=0}=\]

\[-\frac{1}{2}e^{-z}\frac{d}{dz}\Bigg{[}\int_0^zF(\epsilon x) dx+
\int^{\infty}_0e^{-2w}F(\epsilon(w+z))dw
-\int^{\infty}_0e^{-2x}F(\epsilon x)dx \Bigg{]}\Bigg{|}_{z=0} \]
\[=-\frac{1}{2}e^{-z}\Bigg{[}F(0)+\frac{d}{dz}\Big{(}
e^{2z}\int^{\infty}_z e^{-2x}F(\epsilon x)dx\Big{)} \Bigg{]}
\Bigg{|}_{z=0} \]

\[=-\frac{1}{2}e^{-z}\Bigg{[}F(0)+2\int^{\infty}_0 e^{-2x}F(\epsilon x)dx
-F(0) \Bigg{]}\Bigg{|}_{z=0} \]

\[=-e^{-z}\int_0^{\infty} e^{-2x}F(\epsilon x) dx\Bigg{|}_{z=0} \]

\[=-\int_0^{\infty} e^{-x}\Big{(}F(0)+\epsilon\frac{x}{2}F'(0)+
\epsilon^2\frac{1}{2}\left(\frac{x}{2}\right)^2F''(0)+...\Big{)}d
\frac{x}{2}\]

\[
=-\frac{1}{2}F(0)-\frac{1}{4}\epsilon F'(0)+
\frac{1}{8}\epsilon^2F''(0)+...
\]

\noindent
Therefore the final result for the first derivative of $\phi$ at zero is
\[
\phi_y|_{z=0}=-K-K^{-1}\frac{1}{2}\frac{V''|_0}{V_o-C}
-\frac{1}{4}
K^{-2}\left[
\frac{V'''|_0}{V_o-C}-\frac{V''|_0V'|_0}{(V_o-C)^2}\right]+...
\]

\noindent
Applying the boundary condition \ref{linear_rescale_02} at the interface,
\[
KC^2-r[(V_o-C)^2\phi_y|_0-(V_o-c)V'|_0]-G(1-r)=0 ~,
\]
we obtain
\[
KC^2-r[-K(V_o-C)^2-K^{-1}\frac{1}{2}(V_o-C)V''|_0+...
-(V_o-C)V'|_0]-G(1-r)=0 ~.
\]
Scaling $K$ and $C$ so that $K=k/r$ and $C=\sqrt{r}c$, and substituting we
have
\[
kc^2+k(V_o-\sqrt{r}c)^2+r^2k^{-1}\frac{1}{2}(V_o-\sqrt{r}c)V''|_0
+r(V_o-\sqrt{r}c)V'|_0-G+rG=0 ~.
\]
If we expand $c$ in powers of $r^{1/2}$,
\[
c=c_0+c_{1/2}r^{1/2}+c_1r+c_{3/2}r^{3/2}+... ~,
\]
we can obtain each term separately. Here we give only the first few terms:
\[
c_0:~~ kc_0^2+kV_o^2=G \Rightarrow c_o=\sqrt{G/k-V_o^2} ~;
\]
\[
c_{1/2}:~~2kc_oc_{1/2}-2kV_oc_0=0 \Rightarrow c_{1/2}=V_o ~;
\]
\[
c_1:~~kc_{1/2}^2+2kc_0c_1+kc_0^2-2kV_oc_{1/2}+kc_0^2+V_oV'|_0+G=0
\]
\[
\Rightarrow c_1=-\frac{3}{2}c_0-\frac{1}{2}V_oV'|_0/c_0 ~.
\]




\newpage


\newpage


\newpage


\newpage


\newpage


\newpage


\newpage


\newpage
\begin{figure}[htbp]
\epsfysize=2.5in
\epsfxsize=4.0in
\epsfbox{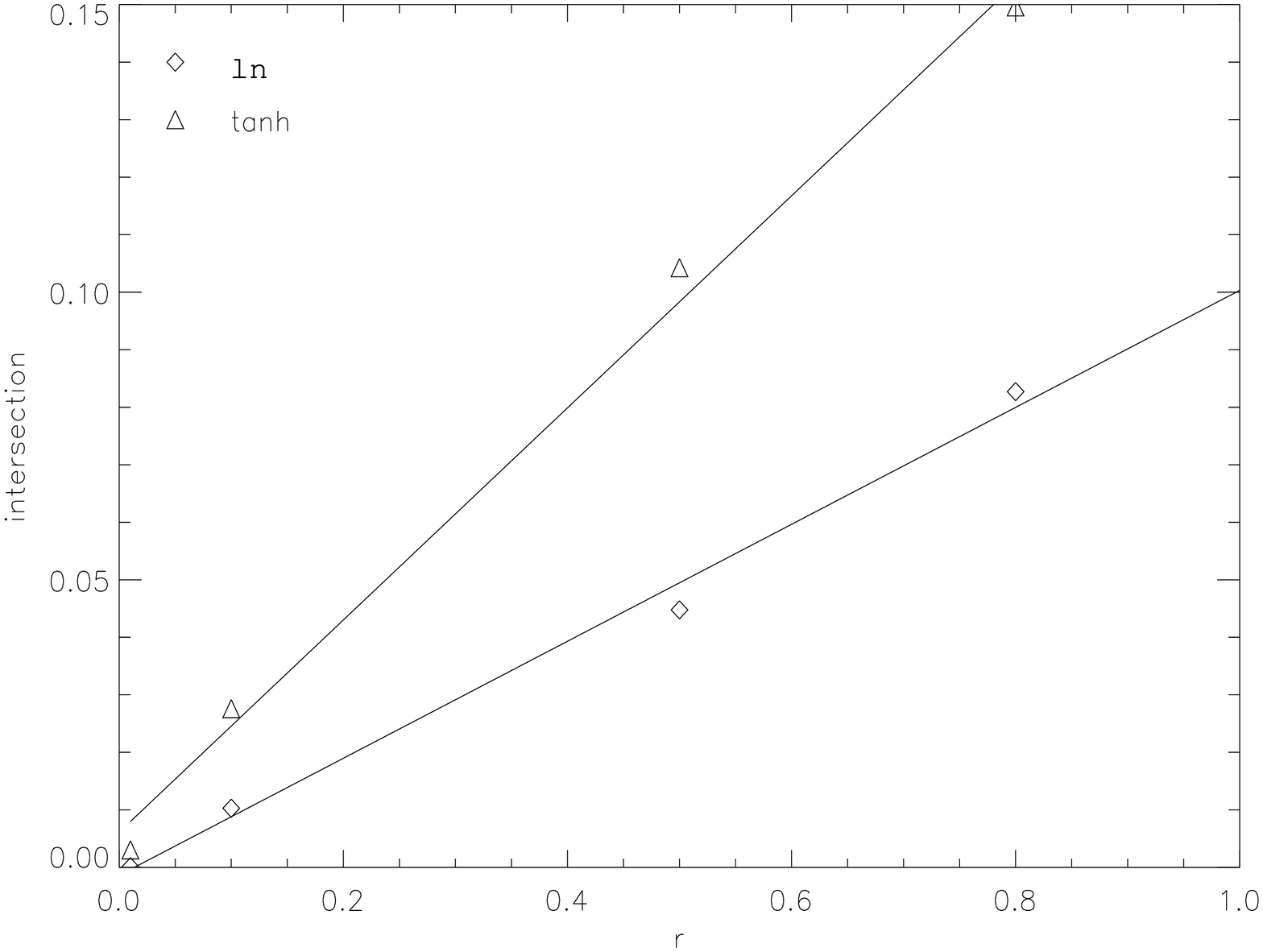}
\caption{Intersections of the previous graph as a function of $r$
(e.g. the dependence of the coefficient in front the exponential
in eq. \ref{empiric}, on $r$).}
\label{fig9}
\end{figure}


\begin{figure}[htbp]
\epsfysize=2.5in
\epsfxsize=4.0in
\epsfbox{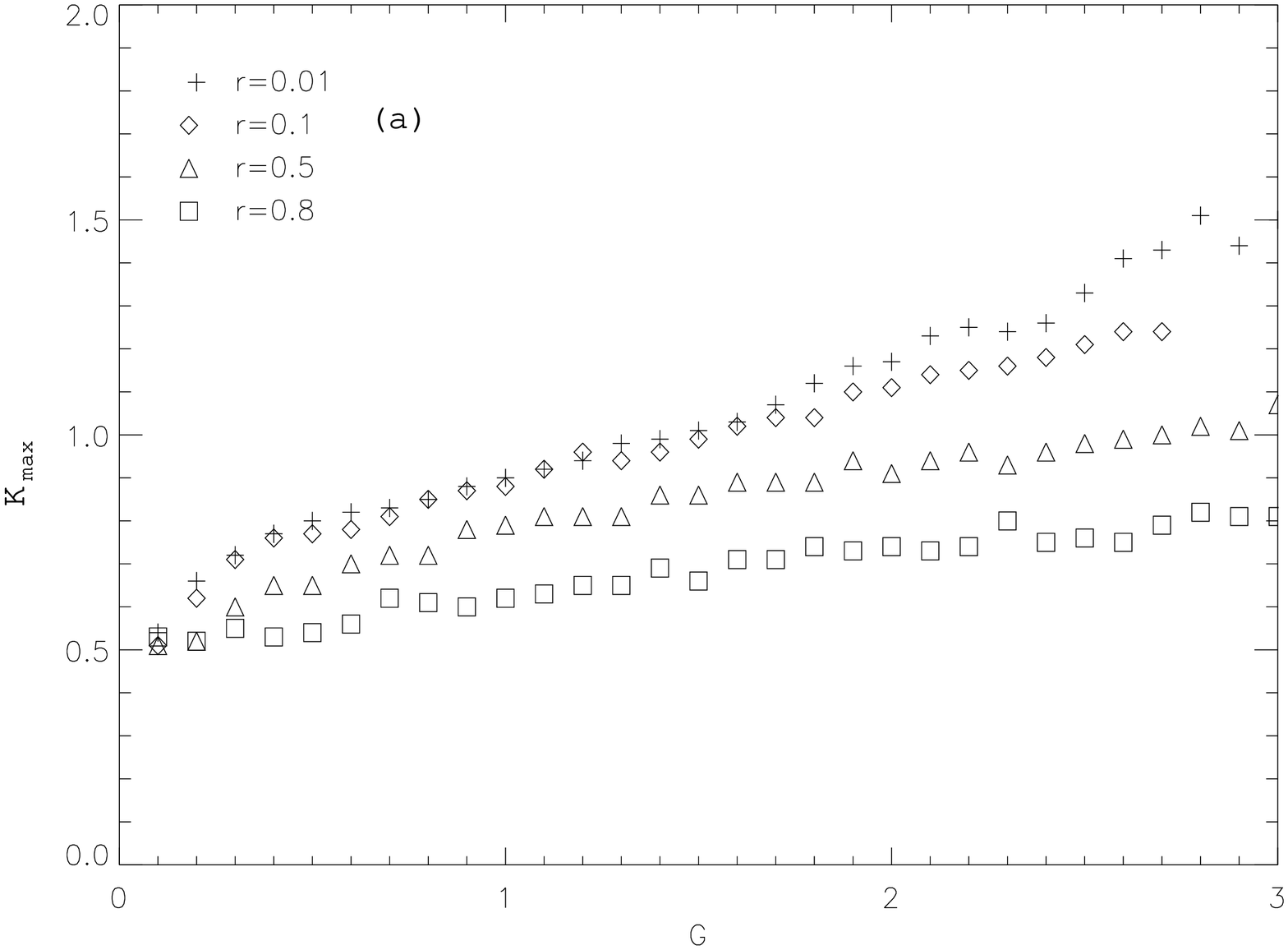}
\epsfysize=2.5in
\epsfxsize=4.0in
\epsfbox{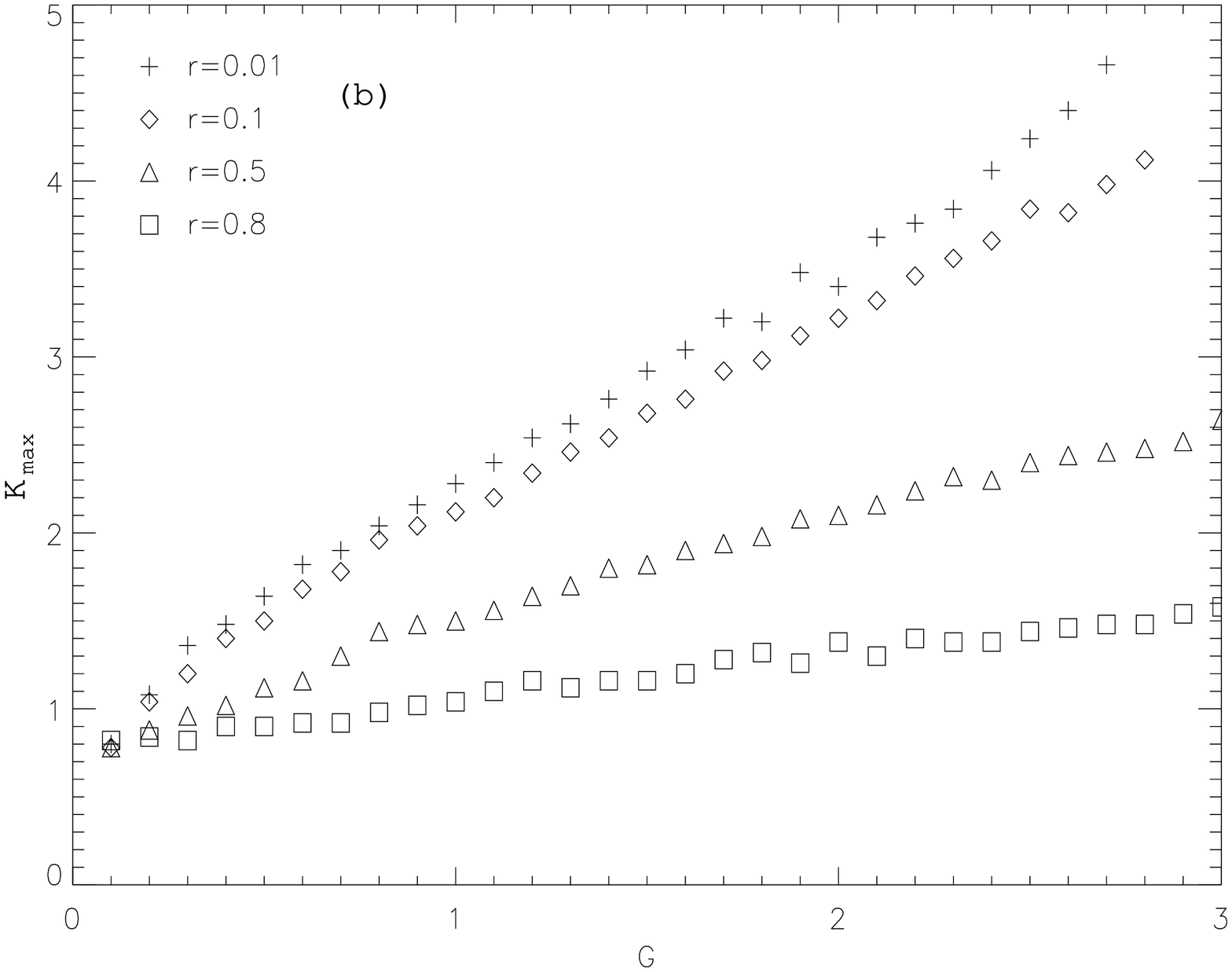}
\caption{The dependence of the wavenumber of the fastest growing mode
on $G$ (a)logarithmic wind profile, (b) tanh wind profile.}
\label{fig10}
\end{figure}


\newpage
\begin{figure}[htbp]
\epsfysize=2.5in
\epsfxsize=4.0in
\epsfbox{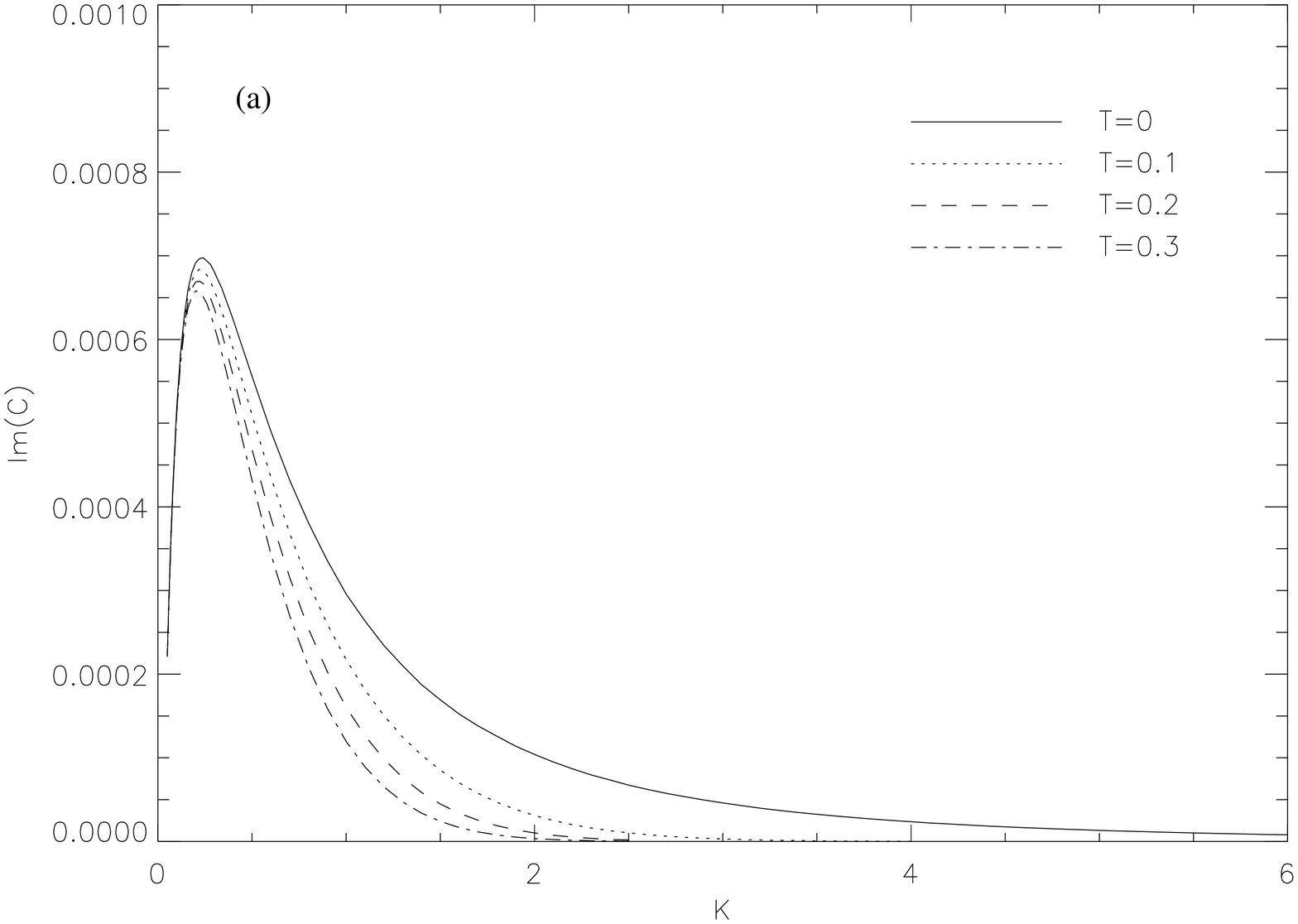}
\epsfysize=2.5in
\epsfxsize=4.0in
\epsfbox{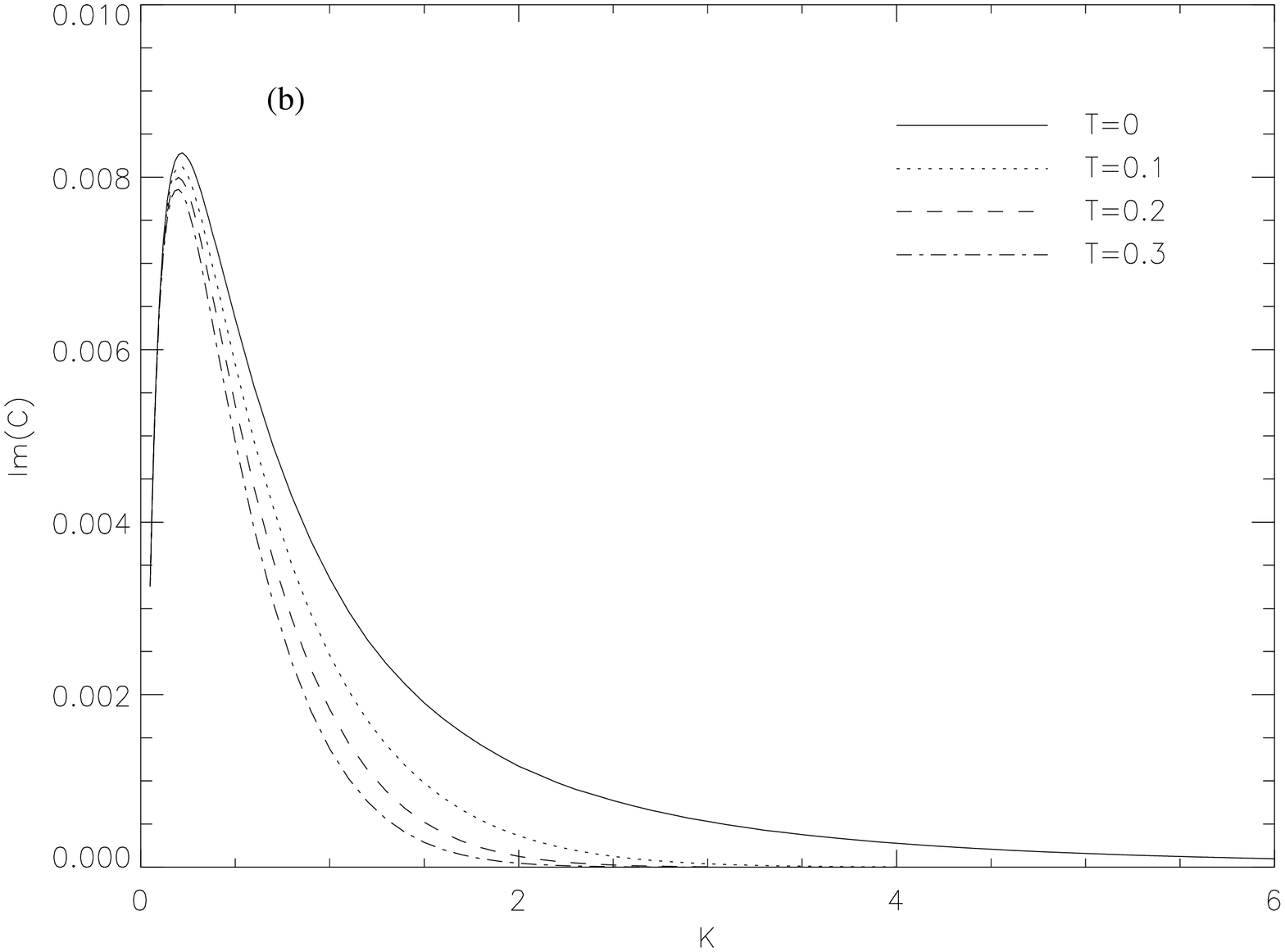}
\epsfysize=2.5in
\epsfxsize=4.0in
\epsfbox{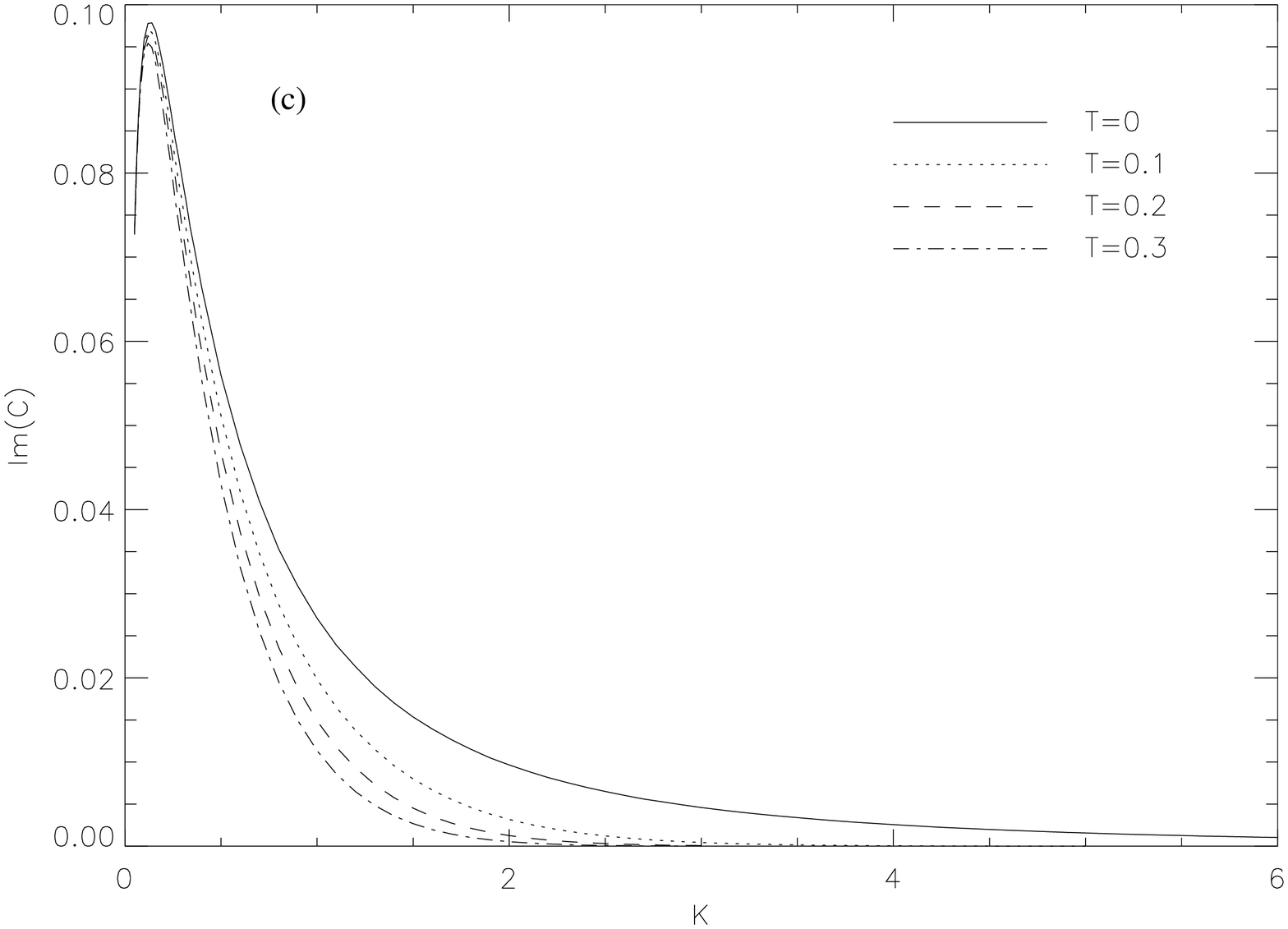}
\caption{Imaginary part of $C$ for a logarithmic  wind profile with G=0.5
in the presence of
surface tension (a)$r=0.01$, (b)$r=0.1$, (c)$r=0.5$ .}
\label{fig11}
\end{figure}


\newpage
\begin{figure}[htbp]
\epsfysize=2.5in
\epsfxsize=4.0in
\epsfbox{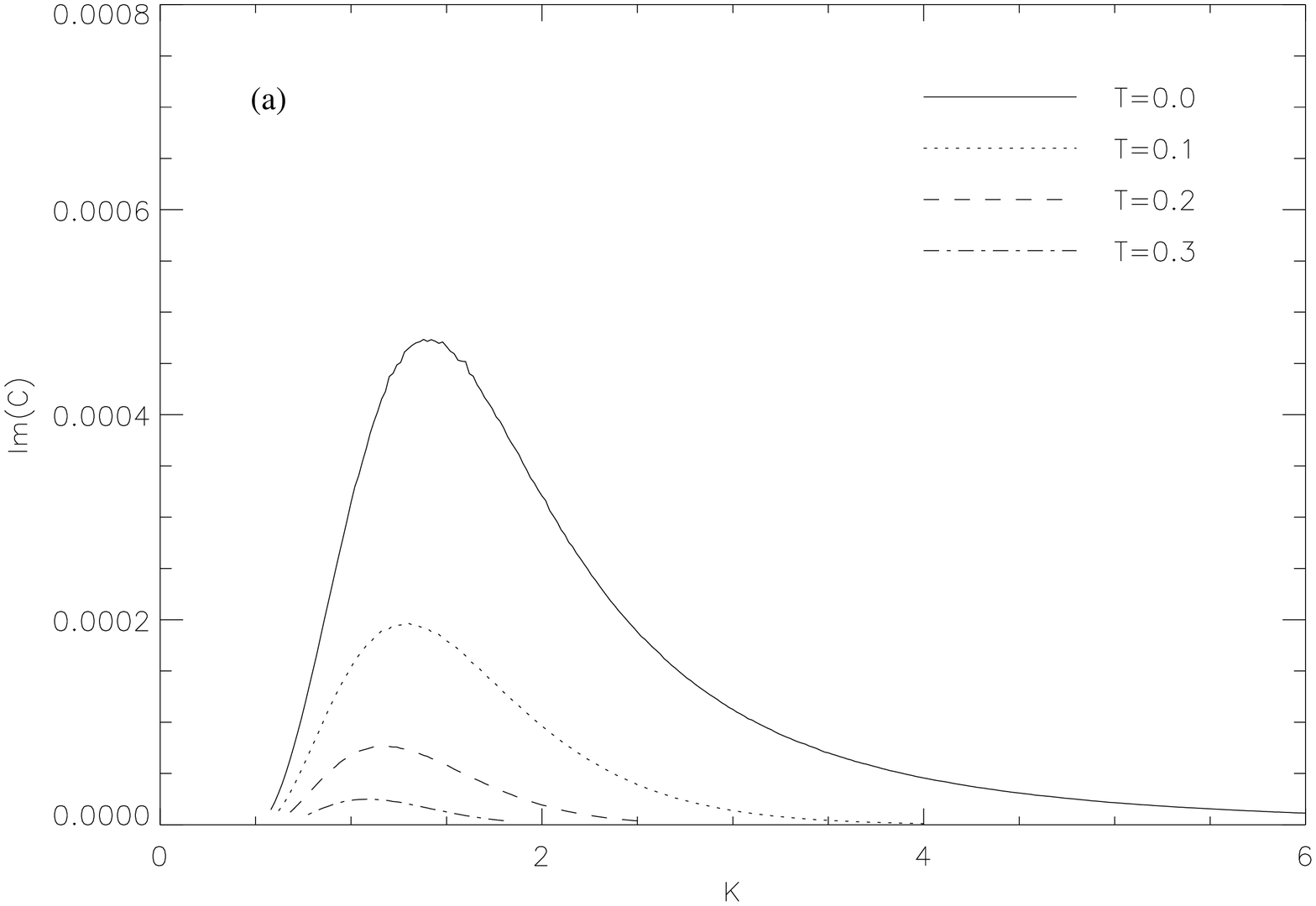}
\epsfysize=2.5in
\epsfxsize=4.0in
\epsfbox{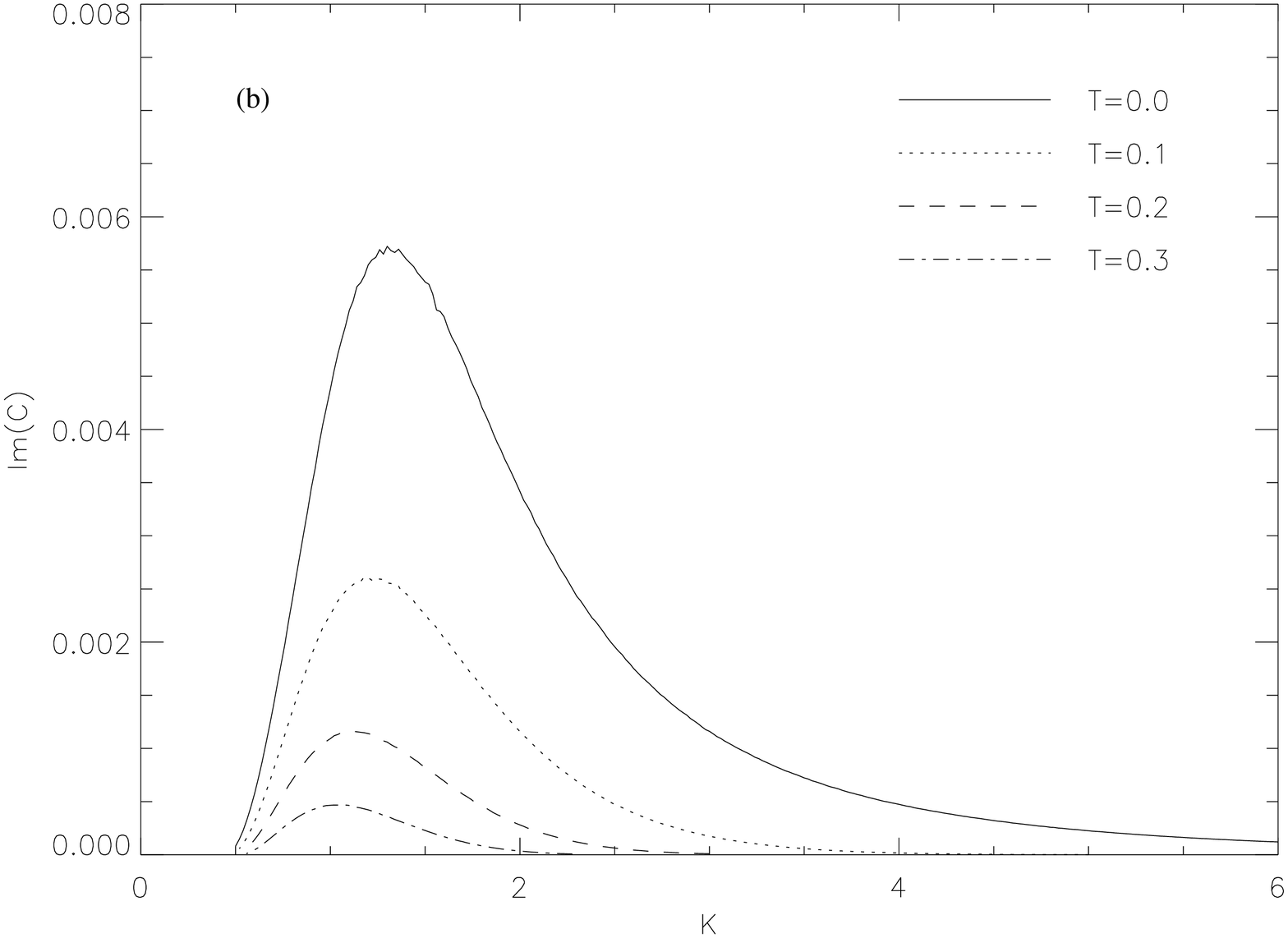}
\epsfysize=2.5in
\epsfxsize=4.0in
\epsfbox{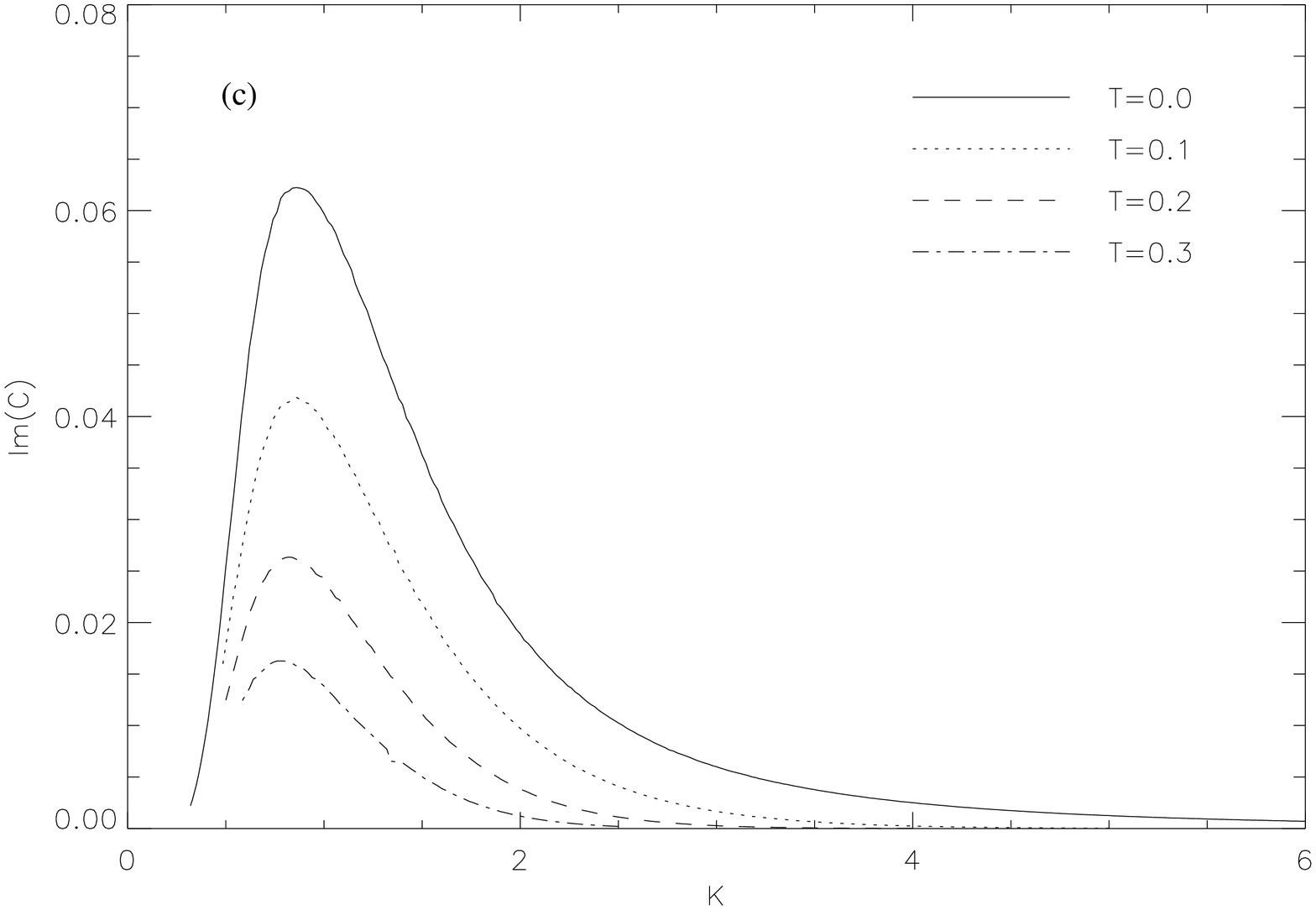}
\caption{Imaginary part of $C$ for a tanh  wind profile with G=0.5
in the presence of
surface tension (a)$r=0.01$, (b)$r=0.1$, (c)$r=0.5$ .}
\label{fig12}
\end{figure}

\newpage


\begin{figure}[htbp]
\epsfysize=2.5in
\epsfxsize=4.0in
\epsfbox{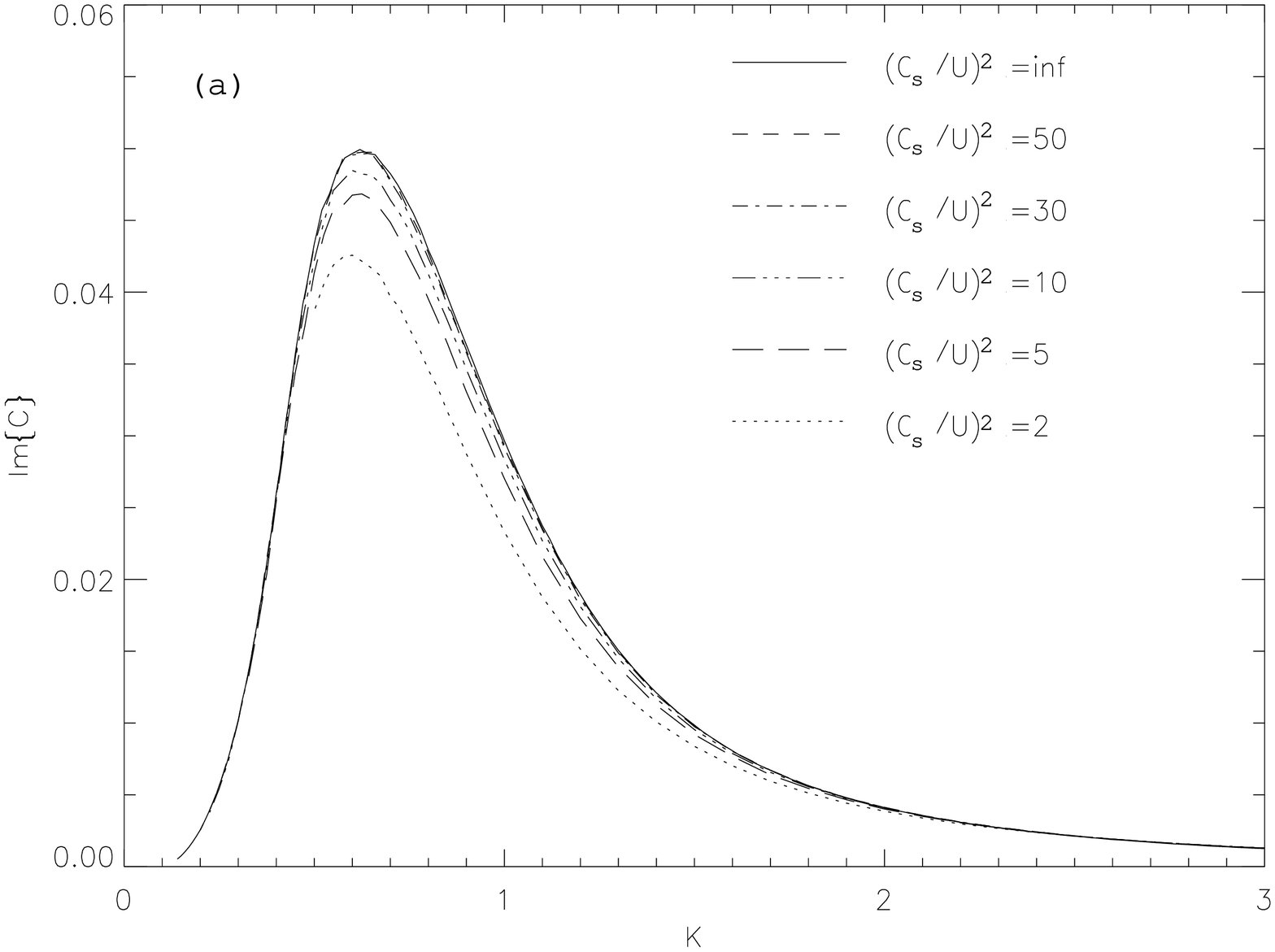}
\epsfysize=2.5in
\epsfxsize=4.0in
\epsfbox{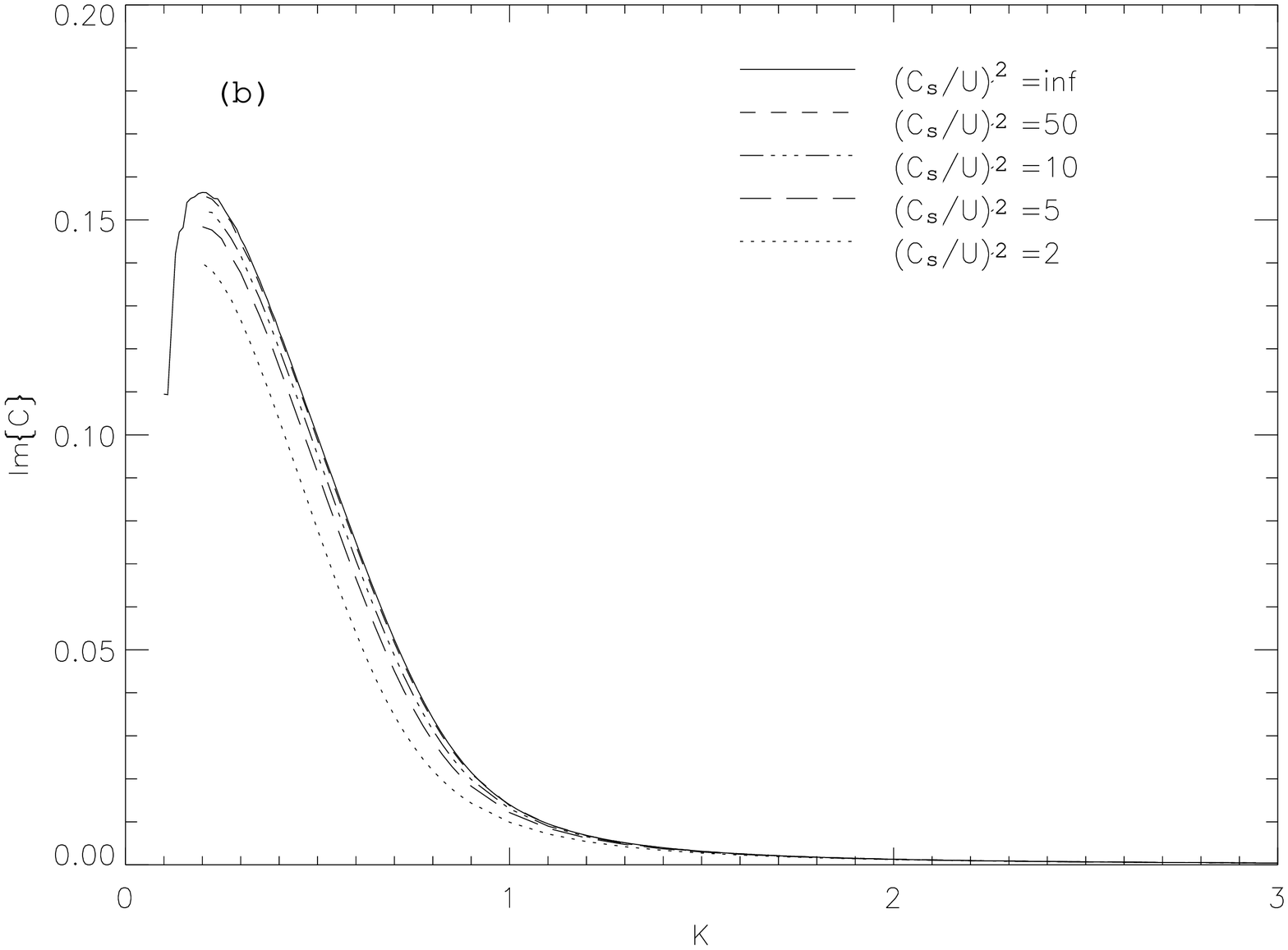}
\caption{Im\{$C$\} for $r$=0.1 for the compressible case
(a)$G$=0.1, (b)$G$=0.01 .}
\label{fig13}
\end{figure}


\begin{table}
\caption{Approximate range for parameter $G$ in three different situations.
\label{table1}}
\begin{tabular}{lcccc}
   & $\su_1$(cm s$^{-1}$) & $g$ cm s$^{-2}$ & $a^{-1}$cm & G\\
\tableline
ocean& $10^2 \sim 10^3$       & $10^3$ 
           & $10\sim 10^2$   & $10^{-1}\sim 1$\\

Sun's surface  & $10^2 \sim 10^5$       & $10^{4.3}$    
           & $10^6 \sim 10^7$            & $10^{1.3}\sim 10^{6.3}$ \\

WD   & $10^4 \sim 10^7$       & $10^8$
           & $10^3 \sim 10^6$            & $1 \sim 10$ \\
\end{tabular}
\end{table}

\end{document}